\documentclass[11pt,a4paper]{article}
\usepackage[T1]{fontenc}
\usepackage{lmodern}
\usepackage[margin=27mm]{geometry}
\usepackage{array,rotating}
\usepackage{amsmath}
\usepackage{amssymb}
\usepackage{amsfonts}
\usepackage{graphicx}
\usepackage{booktabs}
\usepackage{multirow}
\usepackage{amsthm}
\usepackage[authoryear,round]{natbib}
\usepackage[hidelinks]{hyperref}
\hypersetup{
  pdftitle={Noniterative Likelihood-Derived Estimation through Auxiliary Estimating Equations},
  pdfauthor={Pedro L. Ramos; Eduardo Ramos; Francisco A. Rodrigues; Francisco Louzada},
  pdfkeywords={Auxiliary estimating equations, Closed-form estimation, Generalized method of moments, Godambe information, Likelihood-derived estimation, Z-estimation}
}

\numberwithin{equation}{section}
\newtheorem{theorem}{Theorem}[section]
\newtheorem{proposition}[theorem]{Proposition}
\newtheorem{corollary}[theorem]{Corollary}
\theoremstyle{definition}
\newtheorem{definition}[theorem]{Definition}
\theoremstyle{remark}
\newtheorem{remark}[theorem]{Remark}

\newcommand{\bs}{\boldsymbol}
\newcommand{\on}{\operatorname}

\begin{document}

\title{Noniterative Likelihood-Derived Estimation through Auxiliary Estimating Equations}

\author{%
\normalsize Pedro L. Ramos$^{1}$\thanks{Corresponding author. Email: \href{mailto:pedro.ramos@mat.uc.cl}{pedro.ramos@mat.uc.cl}.}
\quad Eduardo Ramos$^{2}$\\[0.3em]
\normalsize Francisco A. Rodrigues$^{2}$ \quad Francisco Louzada$^{2}$\\[0.9em]
\small $^{1}$Facultad de Matem\'aticas, Pontificia Universidad Cat\'olica de Chile,\\
\small Avenida Vicu\~na Mackenna 4860, Macul, Santiago, Chile\\[0.3em]
\small $^{2}$University of S\~ao Paulo, S\~ao Carlos, SP 13566-590, Brazil}
\date{}
\maketitle

\begin{abstract}
Closed-form estimators are useful when parametric models are repeatedly refitted
but likelihood maximization is iterative. We study a likelihood-derived
construction of auxiliary estimating equations obtained by differentiating a
positive auxiliary function and centering the derivatives under a baseline
model. The resulting estimators are just-identified Z-estimators, or
equivalently GMM estimators; the contribution is a constructive route to
explicitly invertible equations rather than a replacement for general
estimating-equation theory. We establish local existence, uniqueness and
asymptotic normality of the selected root, characterize optimal linear
combinations through Godambe information, and give constrained variants and
influence-function diagnostics. Transformed exponential families, Beta,
Weibull AFT regression,
Wishart covariance estimation, zero-inflated counts, tail modelling and copula
dependence illustrate the construction. Simulations and survival-tree split
screening quantify the statistical-computational trade-off.
\end{abstract}


\section{Introduction}

Maximum likelihood estimation is a central principle of parametric inference,
largely because of its likelihood-based interpretation, invariance properties,
and standard large-sample guarantees under regularity conditions
\citep{2004-Bierens, 2006-Lehmann, redner1981note}. In many common models,
however, the likelihood equations have no closed-form solution. Estimation
then relies on numerical optimization, root finding, or iterative algorithms
such as the expectation-maximization algorithm \citep{dempster1977maximum,
ng2013recent}. This is not merely a technical inconvenience: when the same
parametric model must be fitted repeatedly, numerical likelihood maximization
can become the dominant computational cost.

Repeated likelihood fitting appears in simulation, resampling, tree-based
model search, ensemble methods, real-time inference, and embedded statistical
procedures. In these settings, a closed-form estimator can be valuable even
when it is not fully efficient, provided it retains a clear inferential link to
the likelihood. Existing closed-form estimators for specific parametric
families are often derived from model-specific algebraic constructions
\citep{ho2023asymptotically, louzada2019note, nawa2023closed, ramos2020bias,
ye2017closed, zhao2022closed}. We consider the constructive question of how
such equations can be generated from a baseline likelihood and, in favourable
cases, selected to admit an explicit inverse.

Unbiased estimating functions, quasi-likelihood and optimal estimating
equations have a long history \citep{godambe1960optimum, heyde1997quasi}.
Likewise, GMM and Z-estimation provide general consistency and limit theory for
roots of empirical equations \citep{hansen1982large, newey1994large,
1998-Vaart}. Our objective is narrower: starting from a parametric likelihood,
we seek auxiliary derivatives whose centered empirical equations are both
identifying and algebraically invertible. Explicit solvability is therefore a
design constraint on an estimating function, not a new asymptotic paradigm.

We study likelihood-derived auxiliary estimating equations. We use the neutral
term \emph{auxiliary-equation estimator} (AE estimator) for a selected root of
these equations. The ordinary likelihood score is replaced, partially or
completely, by derivatives of a likelihood-type function with respect to
auxiliary coordinates, centered by their model expectations. When the auxiliary
function is the original likelihood, the ordinary likelihood score equations
are recovered. For other
choices, the resulting equations may be explicitly solvable while retaining a
likelihood-derived construction.

Every AE estimator is a just-identified \(Z\)-estimator and, equivalently, a
just-identified GMM estimator. The construction therefore does not replace the
general asymptotic theory for estimating equations. Instead, it specifies a
likelihood-derived representation and gives a direct route from an unbiased
estimating function to an auxiliary function \(g\). Local identification and
the selection of the relevant root remain essential, exactly as in the usual
\(Z\)-estimation theory \citep{1996-Vaart}.

The paper has four contributions. First, it gives a constructive
likelihood-derived representation of unbiased estimating equations through an
exponential tilt and identifies an explicit-inversion criterion based on moment
maps. Second, it records the local conditions under which the resulting root is
well defined and inherits the standard large-sample distribution, including a
constructive measurable-root argument and the independent non-identically
distributed setting relevant to fixed-design regression. Third, it develops
Godambe-optimal combinations, constrained versions and influence-function
diagnostics within the same representation. Fourth, it develops explicit
constructions for transformed exponential families, Beta, Weibull AFT,
matrix-valued Wishart, zero-inflated count, generalized Pareto and copula
models. Monte Carlo experiments and survival-tree split screening quantify the
resulting statistical-computational trade-off.

\section{Likelihood-derived auxiliary estimating equations}\label{sec:theory}

The method proposed here is designed to obtain closed-form estimators for models with a given density \(f(x\,;\,\bs\theta)\), understood with respect to a measure \(\mu\), which may be either discrete or continuous. Let \(\Omega\) denote the sample space, let \(\mathcal X\) denote the space of the observations, and let \(\Theta\subset\mathbb R^s\) be an open parameter space containing the true parameter value \(\bs\theta_0\). For each \(x\in\mathcal X\), let \(\mathcal A_x\subset\mathbb R^r\), \(0\leq r\leq s\), be an open set, possibly depending on \(x\), containing a common fixed value \(\bs\alpha_0\). The vector \(\bs\alpha\) represents auxiliary coordinates used to construct the estimating equations and is fixed at \(\bs\alpha_0\) after differentiation.

Suppose that \(X_1,\ldots,X_n\) are independent and identically distributed random variables with strictly positive density \(f(x\,;\,\bs\theta_0)\). Let \(g(x\,;\,\bs\theta,\bs\alpha)\) be a strictly positive function defined for \(x\in\mathcal X\), \(\bs\theta\in\Theta\), and \(\bs\alpha\in\mathcal A_x\). All expectations \(E_{\bs\theta}\) below are taken with respect to the baseline model \(f(\cdot\,;\,\bs\theta)\). We define the \textit{auxiliary estimating equations for \(\bs\theta\) over the coordinates \((\theta_1,\ldots,\theta_{s-r},\bs\alpha)\) at \(\bs\alpha=\bs\alpha_0\)} as
\begin{equation}\label{modified}
\begin{aligned}
&\frac{1}{n}\sum_{i=1}^n \frac{\partial}{\partial \theta_j}  \log\, g(X_i\,;\,\bs{\theta},\bs{\alpha}_0) =
\on{E}_{\bs{\theta}}\left[\frac{\partial}{\partial \theta_j}  \log\, g(X_1\,;\,\bs{\theta},\bs{\alpha}_0)\right],
\quad 1\leq j\leq s-r,\\
&\frac{1}{n}\sum_{i=1}^n \frac{\partial}{\partial \alpha_j}  \log\, g(X_i\,;\,\bs{\theta},\bs{\alpha}_0)=
\on{E}_{\bs{\theta}}\left[\frac{\partial}{\partial \alpha_j}  \log\, g(X_1\,;\,\bs{\theta},\bs{\alpha}_0)\right],
\quad  1\leq j\leq r,
\end{aligned}
\end{equation}
provided that the partial derivatives exist and the expectations are finite.

These equations contain the ordinary likelihood equations as a special case. Indeed, if the identity
\[
\int_{\mathcal X} f(x\,;\,\bs\theta)\,d\mu=1
\]
can be differentiated under the integral sign, then
\[
\int_{\mathcal X}\frac{\partial}{\partial\theta_j}f(x\,;\,\bs\theta)\,d\mu=0.
\]
Thus, when \(r=0\) and \(g(x\,;\,\bs\theta)=f(x\,;\,\bs\theta)\), the auxiliary estimating equations reduce to
\[
\frac{1}{n}\sum_{i=1}^n
\frac{\partial}{\partial \theta_j}\log f(X_i\,;\,\bs\theta)=0,
\quad 1\leq j\leq s,
\]
which are precisely the ordinary likelihood score equations.

The term AE estimator denotes an estimator obtained from
\eqref{modified}; it does not assert that \(g\) is maximized or that \(g\) is
a probability density. To state the local estimating-equation result, let
\begin{equation}\label{defh}
h_j(x\,;\,\bs{\theta}) =
\frac{\partial}{\partial \beta_j}\log\, g \left(x\,;\,\bs{\theta},\bs{\alpha}_0\right)
-
\on{E}_{\bs{\theta}}\left[
\frac{\partial}{\partial \beta_j}\log\, g \left(X_1\,;\,\bs{\theta},\bs{\alpha}_0\right)
\right],
\end{equation}
for \(1\leq j\leq s\), where
\[
(\beta_1,\ldots,\beta_s)=(\theta_1,\ldots,\theta_{s-r},\alpha_1,\ldots,\alpha_r).
\]
Write \(h=(h_1,\ldots,h_s)^\top\), and define the empirical and population
maps, respectively, by
\[
\Psi_n(\bs\theta)=\frac1n\sum_{i=1}^n h(X_i;\bs\theta),
\qquad
\Psi(\bs\theta)=E_{\bs\theta_0}[h(X_1;\bs\theta)].
\]
Thus \(\Psi(\bs\theta_0)=\bs0\), and \eqref{modified} is precisely
\(\Psi_n(\bs\theta)=\bs0\). Define \(J(\bs\theta)=\left(J_{i,j}(\bs\theta)\right)\in M_s(\mathbb R)\) and \(K(\bs\theta)=\left(K_{i,j}(\bs\theta)\right)\in M_s(\mathbb R)\) by
\begin{equation}\label{eqj}
\begin{aligned}
J_{i,j}(\bs{\theta})&=
\on{E}_{\bs{\theta}}
\left[
-\frac{\partial}{\partial\theta_j}\, h_i(X_1\, ;\, \bs{\theta})
\right],\\
K_{i,j}(\bs{\theta})&=
\on{cov}_{\bs{\theta}}
\left[
h_i(X_1\, ;\, \bs{\theta}),\, h_j(X_1\, ;\, \bs{\theta})
\right],
\end{aligned}
\end{equation}
for \(1\leq i,j\leq s\). In particular,
\(\dot\Psi(\bs\theta_0)=-J(\bs\theta_0)\). These are the usual
sensitivity and variability matrices for an estimating equation.

\begin{theorem}[Local root and standard \(Z\)-estimation limit]\label{coprinc}
Let \(B=\{\bs\theta:\|\bs\theta-\bs\theta_0\|<\delta\}\) and suppose
that its closure \(\overline B\) is contained in \(\Theta\). Assume that
\((\omega,\bs\theta)\mapsto\Psi_n(\omega,\bs\theta)\) is jointly
measurable on the underlying probability space times \(\overline B\), and
suppose that:
\begin{itemize}
\item[(A)] \(\Psi\) is continuously differentiable on \(\overline B\),
\(J_0=J(\bs\theta_0)\) is nonsingular, and
\[
\sup_{\bs\theta\in\overline B}
\left\|I_s+J_0^{-1}\dot\Psi(\bs\theta)\right\|<\frac14.
\]
\item[(B)] With probability tending to one, \(\Psi_n\) is continuously
differentiable on \(\overline B\), and
\[
\sup_{\bs\theta\in\overline B}
\left\|\dot\Psi_n(\bs\theta)-\dot\Psi(\bs\theta)\right\|
\ \longrightarrow_p\ 0.
\]
\item[(C)] \(K_0=K(\bs\theta_0)\) exists and
\[
\sqrt n\,\Psi_n(\bs\theta_0)
\ \overset{D}{\longrightarrow}\ N_s(\bs0,K_0).
\]
\end{itemize}
Then, with probability tending to one, \(\Psi_n(\bs\theta)=\bs0\) has a
unique root \(\hat{\bs\theta}_n\) in \(B\). A measurable version is obtained
by defining it to equal \(\bs\theta_0\) on the exceptional event. It satisfies
\(\hat{\bs\theta}_n\to_p\bs\theta_0\), and
\[
\sqrt n(\hat{\bs\theta}_n-\bs\theta_0)
\ \overset{D}{\longrightarrow}\
N_s\left(\bs0,J_0^{-1}K_0J_0^{-T}\right).
\]
\end{theorem}

\begin{proof}
Put
\[
T_n(\bs\theta)=\bs\theta+J_0^{-1}\Psi_n(\bs\theta).
\]
By (A)--(B), with probability tending to one,
\[
q_n:=\sup_{\bs\theta\in\overline B}
\|\dot T_n(\bs\theta)\|<\frac12.
\]
Condition (C) also implies
\(\|J_0^{-1}\Psi_n(\bs\theta_0)\|=O_p(n^{-1/2})\). Hence, on events whose
probabilities tend to one,
\(\|J_0^{-1}\Psi_n(\bs\theta_0)\|<(1-q_n)\delta\). For every
\(\bs\theta\in\overline B\),
\[
\begin{aligned}
\|T_n(\bs\theta)-\bs\theta_0\|
&\leq
\|T_n(\bs\theta)-T_n(\bs\theta_0)\|
+\|J_0^{-1}\Psi_n(\bs\theta_0)\|\\
&\leq q_n\delta+
\|J_0^{-1}\Psi_n(\bs\theta_0)\|<\delta.
\end{aligned}
\]
Thus \(T_n\) maps the complete set \(\overline B\) into itself and is a
contraction. The Banach fixed-point theorem gives a unique fixed point in
\(\overline B\); the strict inequality above places it in \(B\). This fixed
point is exactly the unique root of \(\Psi_n\) in \(B\).

For measurability, start the Picard iteration at \(\bs\theta_0\) and set
\[
\bs\theta_n^{(m+1)}=T_n(\bs\theta_n^{(m)}).
\]
Joint measurability makes
every iterate measurable, and their pointwise limit on the contraction event
is the fixed point. Defining the root to equal \(\bs\theta_0\) on the
complement gives a measurable version. Moreover,
\[
\|\hat{\bs\theta}_n-\bs\theta_0\|
\leq
\frac{\|J_0^{-1}\Psi_n(\bs\theta_0)\|}{1-q_n}
=O_p(n^{-1/2}),
\]
which proves consistency. Finally, the integral Taylor expansion gives
\[
\bs0=\Psi_n(\bs\theta_0)+
\left\{\int_0^1
\dot\Psi_n\bigl(\bs\theta_0+t(\hat{\bs\theta}_n-\bs\theta_0)\bigr)
\,dt\right\}
(\hat{\bs\theta}_n-\bs\theta_0).
\]
The matrix in braces converges in probability to
\(\dot\Psi(\bs\theta_0)=-J_0\). Condition (C), Slutsky's theorem and the
preceding expansion yield the stated normal limit. This is the standard local
\(Z\)-estimation limit, with the contraction argument supplying the local root
and its selection \citep{1998-Vaart, newey1994large}.
\end{proof}

Assumption (A) is a local identification condition. It deliberately makes no
claim about roots outside \(B\). If an algebraic construction has several
admissible roots, its estimator must state which branch is selected; the theorem
validates the branch that lies in the locally identified neighbourhood.

Exponential-type auxiliary functions are convenient because their derivatives
are linear combinations of observable statistics. They provide an easy route to
checking the hypotheses of Theorem~\ref{coprinc}, but do not require a separate
asymptotic theorem. In particular, \(g\) need not be a probability density:
the baseline data-generating model remains \(f\), whereas \(g\) only generates
the estimating function \(h\).

Several relevant applications, especially regression models with fixed
covariates, involve independent but non-identically distributed observations.
For completeness, the following standard extension records the conditions used
for that setting; it assumes a consistent selected root.

\begin{proposition}[Independent observations]
Let \(Z_1,\ldots,Z_n\) be independent random variables, not necessarily
identically distributed. For each \(i\), let \(g_i\) be a likelihood-type
function, possibly depending on fixed covariates, and define
\[
\psi_i(z;\bs\theta)
=
\frac{\partial}{\partial\bs\beta}
\log g_i(z;\bs\theta,\bs\alpha_0)
-
E_{\bs\theta}
\left[
\frac{\partial}{\partial\bs\beta}
\log g_i(Z_i;\bs\theta,\bs\alpha_0)
\right],
\]
where the derivative with respect to \(\bs\beta\) is taken over the same
\(s\) coordinates used in the auxiliary estimating equations. Suppose that
\(E_{\bs\theta_0}[\psi_i(Z_i;\bs\theta_0)]=0\),
\[
A_n(\bs\theta_0)
=
\frac1n\sum_{i=1}^n
E_{\bs\theta_0}
\left[
\frac{\partial}{\partial\bs\theta^\top}
\psi_i(Z_i;\bs\theta_0)
\right]
\to A(\bs\theta_0),
\]
where \(A(\bs\theta_0)\) is nonsingular, and
\[
B_n(\bs\theta_0)
=
\frac1n\sum_{i=1}^n
\operatorname{Var}_{\bs\theta_0}
\left[
\psi_i(Z_i;\bs\theta_0)
\right]
\to B(\bs\theta_0).
\]
Assume also the usual uniform law of large numbers and central limit theorem
conditions for \(\psi_i\) and its derivatives. Then any consistent solution
\(\hat{\bs\theta}_n\) of
\[
\sum_{i=1}^n \psi_i(Z_i;\bs\theta)=0
\]
satisfies
\[
\sqrt n(\hat{\bs\theta}_n-\bs\theta_0)
\overset{D}{\to}
N_s\left(
0,
A(\bs\theta_0)^{-1}
B(\bs\theta_0)
A(\bs\theta_0)^{-T}
\right).
\]
\end{proposition}

\begin{proof}
Let \(\Psi_n(\bs\theta)=n^{-1}\sum_{i=1}^n\psi_i(Z_i;\bs\theta)\). By
assumption, \(\Psi_n(\hat{\bs\theta}_n)=0\), \(\hat{\bs\theta}_n\to
\bs\theta_0\) in probability, and the derivative
\(\partial\Psi_n(\bs\theta)/\partial\bs\theta^\top\) converges uniformly in a
neighborhood of \(\bs\theta_0\) to \(A(\bs\theta_0)\). A Taylor expansion gives
\[
0
=
\Psi_n(\bs\theta_0)
+
\left\{
\frac{\partial}{\partial\bs\theta^\top}
\Psi_n(\tilde{\bs\theta}_n)
\right\}
(\hat{\bs\theta}_n-\bs\theta_0),
\]
where \(\tilde{\bs\theta}_n\) lies between \(\hat{\bs\theta}_n\) and
\(\bs\theta_0\). The central limit theorem for independent non-identically
distributed arrays gives
\[
\sqrt n\,\Psi_n(\bs\theta_0)
\overset{D}{\to}
N_s(0,B(\bs\theta_0)).
\]
Combining these two facts with Slutsky's theorem yields the stated covariance
matrix. The minus sign from the Taylor expansion does not affect the covariance.
\end{proof}

We next discuss invariance under one-to-one transformations. Let
\(\pi:\Theta\to\Lambda\) be a diffeomorphism and define
\[
g^*(x\,;\,\bs\lambda,\bs\alpha)
=
g(x\,;\,\pi^{-1}(\bs\lambda),\bs\alpha),
\qquad \bs\lambda\in\Lambda.
\]
The auxiliary estimating equations for
\(\bs\lambda=\pi(\bs\theta)\), evaluated at \(\bs\alpha=\bs\alpha_0\), are
defined as
\begin{equation}\label{modified2}
\begin{aligned}
&\frac{1}{n}\sum_{i=1}^n
\frac{\partial}{\partial \lambda_j}
\log g^*(X_i\,;\,\bs{\lambda},\bs{\alpha}_0)
=
\on{E}_{\pi^{-1}(\bs{\lambda})}
\left[
\frac{\partial}{\partial \lambda_j}
\log g^*(X_1\,;\,\bs{\lambda},\bs{\alpha}_0)
\right],\\
&\frac{1}{n}\sum_{i=1}^n
\frac{\partial}{\partial \alpha_j}
\log g^*(X_i\,;\,\bs{\lambda},\bs{\alpha}_0)
=
\on{E}_{\pi^{-1}(\bs{\lambda})}
\left[
\frac{\partial}{\partial \alpha_j}
\log g^*(X_1\,;\,\bs{\lambda},\bs{\alpha}_0)
\right].
\end{aligned}
\end{equation}
The first line applies for \(1\leq j\leq s-r\), and the second for
\(1\leq j\leq r\), provided that the partial derivatives exist and the
expectations are finite.

\begin{proposition}[One-to-one invariance]
Suppose \(\Theta=\Theta_1\times\Theta_2\) and
\(\Lambda=\Lambda_1\times\Lambda_2\), where
\(\Theta_1,\Lambda_1\subset\mathbb R^{s-r}\) and
\(\Theta_2,\Lambda_2\subset\mathbb R^r\) are open sets. Suppose that
\(\pi:\Theta\to\Lambda\) can be written as
\[
\pi(\bs\theta)
=
(\pi_1(\bs\theta_1),\pi_2(\bs\theta_2)),
\qquad
\bs\theta=(\bs\theta_1,\bs\theta_2)\in\Theta_1\times\Theta_2,
\]
where \(\pi_1:\Theta_1\to\Lambda_1\) and
\(\pi_2:\Theta_2\to\Lambda_2\) are diffeomorphisms. Suppose that, for some
\(n\in\mathbb N\), \(\hat{\bs\theta}_n(\bs X)\) is, with probability one, a
solution of the auxiliary estimating equations for \(\bs\theta\) at
\(\bs\alpha=\bs\alpha_0\). Then, with probability one,
\[
\hat{\bs\lambda}_n(\bs X)
=
\pi(\hat{\bs\theta}_n(\bs X))
\]
is a solution of the auxiliary estimating equations for
\(\bs\lambda=\pi(\bs\theta)\) at \(\bs\alpha=\bs\alpha_0\).
\end{proposition}

\begin{proof}
Since \(\pi\) is block-separable, we have
\[
\pi^{-1}(\bs\lambda)
=
(\pi_1^{-1}(\bs\lambda_1),\pi_2^{-1}(\bs\lambda_2)).
\]
Let
\[
\pi_1^{-1}(\bs\lambda_1)
=
(\pi^*_{11}(\bs\lambda_1),\ldots,\pi^*_{1(s-r)}(\bs\lambda_1)).
\]
By the chain rule,
\begin{equation}\label{prop12}
\frac{\partial}{\partial \lambda_j}
\log g^*(X_i\,;\,\bs\lambda,\bs\alpha_0)
=
\sum_{k=1}^{s-r}
\frac{\partial}{\partial \theta_k}
\log g(X_i\,;\,\pi^{-1}(\bs\lambda),\bs\alpha_0)
\frac{\partial}{\partial \lambda_j}
\pi^*_{1k}(\bs\lambda_1),
\end{equation}
for \(1\leq j\leq s-r\). By hypothesis, with probability one,
\(\hat{\bs\theta}_n(\bs X)\) satisfies
\begin{equation}\label{prop11}
\frac{1}{n}\sum_{i=1}^n
\frac{\partial}{\partial \theta_j}
\log g(X_i\,;\,\hat{\bs\theta}_n(\bs X),\bs\alpha_0)
=
E_{\hat{\bs\theta}_n(\bs X)}
\left[
\frac{\partial}{\partial \theta_j}
\log g(X_1\,;\,\hat{\bs\theta}_n(\bs X),\bs\alpha_0)
\right],
\end{equation}
for \(1\leq j\leq s-r\). Therefore, using \eqref{prop12} and
\eqref{prop11}, with
\(\hat{\bs\lambda}_n(\bs X)=\pi(\hat{\bs\theta}_n(\bs X))\), we obtain
\[
\begin{aligned}
&\frac{1}{n}\sum_{i=1}^n
\frac{\partial}{\partial \lambda_j}
\log g^*(X_i\,;\,\hat{\bs\lambda}_n(\bs X),\bs\alpha_0)\\
&=
\sum_{k=1}^{s-r}
\left\{
\frac{1}{n}\sum_{i=1}^n
\frac{\partial}{\partial \theta_k}
\log g(X_i\,;\,\hat{\bs\theta}_n(\bs X),\bs\alpha_0)
\right\}
\frac{\partial}{\partial \lambda_j}
\pi^*_{1k}(\hat{\bs\lambda}_{1n}(\bs X))\\
&=
\sum_{k=1}^{s-r}
E_{\hat{\bs\theta}_n(\bs X)}
\left[
\frac{\partial}{\partial \theta_k}
\log g(X_1\,;\,\hat{\bs\theta}_n(\bs X),\bs\alpha_0)
\right]
\frac{\partial}{\partial \lambda_j}
\pi^*_{1k}(\hat{\bs\lambda}_{1n}(\bs X))\\
&=
E_{\pi^{-1}(\hat{\bs\lambda}_n(\bs X))}
\left[
\frac{\partial}{\partial \lambda_j}
\log g^*(X_1\,;\,\hat{\bs\lambda}_n(\bs X),\bs\alpha_0)
\right].
\end{aligned}
\]
Hence \(\hat{\bs\lambda}_n(\bs X)\) satisfies the first set of equations in
\eqref{modified2}.

For the auxiliary coordinates, since \(\pi^{-1}(\bs\lambda)\) does not depend
on \(\bs\alpha\),
\[
\frac{\partial}{\partial \alpha_j}
\log g^*(X_i\,;\,\hat{\bs\lambda}_n(\bs X),\bs\alpha_0)
=
\frac{\partial}{\partial \alpha_j}
\log g(X_i\,;\,\hat{\bs\theta}_n(\bs X),\bs\alpha_0).
\]
Using again that \(\hat{\bs\theta}_n(\bs X)\) satisfies the auxiliary
likelihood equations, it follows that
\[
\begin{aligned}
&\frac{1}{n}\sum_{i=1}^n
\frac{\partial}{\partial \alpha_j}
\log g^*(X_i\,;\,\hat{\bs\lambda}_n(\bs X),\bs\alpha_0)
\\
&\qquad=
E_{\pi^{-1}(\hat{\bs\lambda}_n(\bs X))}
\left[
\frac{\partial}{\partial \alpha_j}
\log g^*(X_1\,;\,\hat{\bs\lambda}_n(\bs X),\bs\alpha_0)
\right].
\end{aligned}
\]
for \(1\leq j\leq r\). Thus \(\hat{\bs\lambda}_n(\bs X)\) satisfies
\eqref{modified2}, completing the proof.
\end{proof}

In general, AE estimators need not be functions of sufficient statistics.
In some applications, \(g\) is chosen as a generalized version of the baseline
density \(f\). In others, \(g\) may be only a likelihood-type function designed
to generate useful estimating equations. This flexibility is one of the main
features of the proposed approach: the function \(g\) is not part of the data
generating model unless explicitly specified, but rather a device for producing
likelihood-based equations that may lead to closed-form estimators.

Every solution of \eqref{modified} is a just-identified \(Z\)-estimator and
can also be viewed as a just-identified GMM estimator. The usual
estimating-equation theory therefore supplies the large-sample result once
local identification, a local root, and the required uniform convergence are
verified \citep{hansen1982large, newey1994large, 1998-Vaart}. Multiple roots
do not remove the need for identification: they make the selection rule part of
the definition of the estimator. Theorem~\ref{coprinc} is intentionally local
and only certifies the root in the identified neighbourhood.

\subsection{A constructive approach for choosing g}

A natural question arising from the proposed method is how to choose the
function \(g\). We now provide a constructive answer. The idea is to start
from unbiased estimating functions and then construct a likelihood-type
function \(g\) whose auxiliary estimating equations reproduce the corresponding
estimating equations.

\begin{definition}
Let \(U:\mathcal X\times\Theta\to\mathbb R^s\) be a measurable function,
\(U=(U_1,\ldots,U_s)\), such that the following expectation is finite. We say
that \(U\) is an unbiased estimating function for \(\bs\theta\) if
\[
E_{\bs\theta}\left[U(X;\bs\theta)\right]=0
\]
for every \(\bs\theta\in\Theta\).
\end{definition}

\begin{proposition}[Auxiliary representation]
Let \(f(x;\bs\theta)\) be a strictly positive density and let
\(U:\mathcal X\times\Theta\to\mathbb R^s\) be an unbiased estimating
function. Define
\[
g_U(x;\bs\theta,\bs a)
=
f(x;\bs\theta)
\exp\left\{\bs a^\top U(x;\bs\theta)\right\},
\qquad \bs a\in\mathbb R^s.
\]
Then, taking the auxiliary coordinates to be \(\bs a\) and evaluating at
\(\bs a_0=\bs0\), the auxiliary estimating equations are
\[
\frac1n\sum_{i=1}^n U_j(X_i;\bs\theta)=0,
\qquad j=1,\ldots,s.
\]
\end{proposition}

\begin{proof}
Since
\[
\log g_U(x;\bs\theta,\bs a)
=
\log f(x;\bs\theta)+\bs a^\top U(x;\bs\theta),
\]
we have
\[
\left.
\frac{\partial}{\partial a_j}
\log g_U(x;\bs\theta,\bs a)
\right|_{\bs a=\bs0}
=
U_j(x;\bs\theta).
\]
Therefore, the auxiliary likelihood equation associated with \(a_j\) is
\[
\frac1n\sum_{i=1}^n U_j(X_i;\bs\theta)
=
E_{\bs\theta}\left[U_j(X;\bs\theta)\right].
\]
Since \(U\) is unbiased, the expectation on the
right-hand side is zero, and the result follows.
\end{proof}

This representation is deliberately inclusive: it shows that the auxiliary
construction is a likelihood-derived way to encode an existing unbiased
estimating equation, not a class larger than the usual estimating-equation
framework.

\begin{corollary}[Moment generator]
Let \(T:\mathcal X\to\mathbb R^s\) be a measurable function such that
\[
m(\bs\theta)=E_{\bs\theta}[T(X)]
\]
is finite for every \(\bs\theta\in\Theta\). Suppose that
\(m:\Theta\to m(\Theta)\) is one-to-one. Define
\[
U_T(x;\bs\theta)=T(x)-m(\bs\theta)
\]
and
\[
g_T(x;\bs\theta,\bs a)
=
f(x;\bs\theta)
\exp\left\{\bs a^\top\left(T(x)-m(\bs\theta)\right)\right\}.
\]
Then the auxiliary estimating equations over the auxiliary coordinates
\(\bs a\), evaluated at \(\bs a_0=\bs0\), are
\[
\frac1n\sum_{i=1}^n T(X_i)=m(\bs\theta).
\]
Whenever \(n^{-1}\sum_{i=1}^nT(X_i)\in m(\Theta)\), if \(m^{-1}\) is
available in closed form, then the corresponding estimator is
\[
\hat{\bs\theta}_{\mathrm{AE}}
=
m^{-1}\left(
\frac1n\sum_{i=1}^n T(X_i)
\right).
\]
\end{corollary}

\begin{proof}
Since
\[
E_{\bs\theta}[U_T(X;\bs\theta)]
=
E_{\bs\theta}[T(X)]-m(\bs\theta)=0,
\]
\(U_T\) is an unbiased estimating function. The result follows directly
from the previous proposition.
\end{proof}

The moment construction also gives a direct rank criterion and a covariance
formula that can be checked before deriving any auxiliary density.

\begin{proposition}[Local moment inversion]\label{prop:moment-limit}
Let \(T:\mathcal X\to\mathbb R^s\) have finite second moments at
\(\bs\theta_0\), and put
\[
m(\bs\theta)=E_{\bs\theta}[T(X)],\qquad
\Omega_0=\operatorname{Var}_{\bs\theta_0}[T(X)].
\]
Suppose that \(m\) is continuously differentiable near \(\bs\theta_0\) and
that \(M_0=\dot m(\bs\theta_0)\) is nonsingular. Then there are
neighbourhoods \(N\) of \(\bs\theta_0\) and \(V\) of
\(m(\bs\theta_0)\) such that \(m:N\to V\) has a continuously
differentiable inverse. With probability tending to one,
\(\bar T_n=n^{-1}\sum_iT(X_i)\) belongs to \(V\), and the selected local
root
\(\hat{\bs\theta}_n=m^{-1}(\bar T_n)\) satisfies
\[
\sqrt n(\hat{\bs\theta}_n-\bs\theta_0)
\overset{D}{\longrightarrow}
N_s\left(\bs0,M_0^{-1}\Omega_0M_0^{-T}\right).
\]
\end{proposition}

\begin{proof}
The inverse function theorem gives \(N\), \(V\) and the differentiable local
inverse, whose derivative at \(m(\bs\theta_0)\) is \(M_0^{-1}\). The law of
large numbers places \(\bar T_n\) in \(V\) with probability tending to one.
The multivariate central limit theorem and delta method then give the stated
limit. Nonsingularity of \(M_0\) is also the local identification condition,
because the population estimating map is
\(\Psi(\bs\theta)=m(\bs\theta_0)-m(\bs\theta)\).
\end{proof}

This result suggests a reproducible way to find \(g\). One first lists
statistics whose expectations are available under the baseline model,
including transformations, factorial moments, matrix moments or derivatives
with respect to an embedding parameter. Candidate statistics are screened by
the rank of their expectation map. Among full-rank subsets, one retains those
for which the empirical equations can be inverted algebraically and states the
admissible branch explicitly. The exponential representation then produces a
positive \(g\); it is the centered estimating function, rather than the
normalizing constant of \(g\), that determines the estimator. Finally,
Godambe information, influence functions and finite-sample admissibility can be
used to compare the remaining explicit candidates. Thus choosing \(g\) is not
an unrestricted search over positive functions: it is a sequence of moment
selection, rank verification, algebraic inversion and statistical comparison.

\begin{remark}
The previous result transforms the problem of choosing \(g\) into the problem
of finding suitable estimating functions. In particular, if one can find
statistics \(T_1(X),\ldots,T_s(X)\) whose expectations are known functions of
\(\bs\theta\), and if the map
\[
\bs\theta\mapsto
\left(E_{\bs\theta}[T_1(X)],\ldots,E_{\bs\theta}[T_s(X)]\right)
\]
can be inverted explicitly, then the corresponding estimator has a closed-form
expression on the event that the empirical moment belongs to \(m(\Theta)\).
When \(m(\bs\theta_0)\) is an interior point of \(m(\Theta)\) and the law of
large numbers applies, that event has probability tending to one.
\end{remark}

\begin{remark}
The functions \(g_U\) and \(g_T\) are not required to be probability density
functions. This is consistent with the general auxiliary-equation
construction, where \(g\) is used to generate
likelihood-based estimating equations and need not be the data-generating
density. If a normalized version is desired and the moment generating function
of \(T(X)\) exists in a neighborhood of the origin, one may instead define
\[
g_T^*(x;\bs\theta,\bs a)
=
f(x;\bs\theta)
\exp\left\{
\bs a^\top T(x)-A(\bs\theta,\bs a)
\right\},
\]
where
\[
A(\bs\theta,\bs a)
=
\log E_{\bs\theta}
\left[\exp\{\bs a^\top T(X)\}\right].
\]
Then \(g_T^*\) is a probability density function and, at \(\bs a=\bs0\),
leads to the same auxiliary estimating equations as those obtained from
\(g_T\).
\end{remark}

Some closed-form estimators obtained from auxiliary estimating equations are
initially defined as unconstrained solutions. For finite samples, these
solutions may occasionally fall outside the admissible parameter space,
especially in models with bounded parameters or support restrictions, such as
zero-inflated models, tail-index models, and copula models. A simple way to
handle this issue is to project the unconstrained estimator onto the admissible
parameter set. When the admissible set is closed and convex, its Euclidean
projection is single valued. Proposition~\ref{prop:projection} shows that this
operation does not affect first-order asymptotic behaviour when the true
parameter is an interior point.

\subsection{Constrained estimators and efficiency}

The unrestricted auxiliary equations may occasionally produce roots outside an
admissible parameter space. The following elementary result states precisely
when a projection used for finite-sample stabilization is asymptotically
irrelevant.

\begin{proposition}[Constrained version]\label{prop:projection}
Let \(\tilde{\bs\theta}_n\) be an unconstrained AE estimator and let
\(\Theta_c\subset\mathbb R^s\) be a closed convex admissible set. Define
\[
\hat{\bs\theta}_n=\Pi_{\Theta_c}(\tilde{\bs\theta}_n),
\]
where \(\Pi_{\Theta_c}\) is the Euclidean projection. If
\(\bs\theta_0\in\operatorname{int}(\Theta_c)\) and
\(\tilde{\bs\theta}_n\to_p\bs\theta_0\), then
\[
P(\hat{\bs\theta}_n=\tilde{\bs\theta}_n)\longrightarrow1.
\]
Consequently, the two estimators have the same first-order asymptotic
distribution.
\end{proposition}

\begin{proof}
Choose \(\epsilon>0\) such that the open ball of radius \(\epsilon\) about
\(\bs\theta_0\) is contained in \(\Theta_c\). Consistency implies that
\(P(\|\tilde{\bs\theta}_n-\bs\theta_0\|<\epsilon)\to1\). On this event the
unconstrained root already belongs to \(\Theta_c\), and its projection is
itself.
\end{proof}

The interior condition is essential. At a boundary point, projection can
change the limiting law and may produce a cone projection or a distribution
with point mass on the boundary. In applications below we therefore project
onto a specified closed inner approximation to an open parameter space, rather
than referring to a projection onto an open set.

The constructive representation also allows a transparent comparison among
several candidate equations. Let
\(U:\mathcal X\times\Theta\to\mathbb R^q\), \(q\geq s\), satisfy
\(E_{\bs\theta}[U(X;\bs\theta)]=0\). For
\(W(\bs\theta)\in\mathbb R^{s\times q}\), put
\[
\psi_W(x;\bs\theta)=W(\bs\theta)U(x;\bs\theta)
\]
and define
\[
g_W(x;\bs\theta,\bs a)
=f(x;\bs\theta)
\exp\{\bs a^\top\psi_W(x;\bs\theta)\},
\qquad \bs a\in\mathbb R^s.
\]
The auxiliary equations over \(\bs a\), evaluated at \(\bs a=\bs0\), are
\(n^{-1}\sum_i\psi_W(X_i;\bs\theta)=\bs0\).

\begin{theorem}[Linear combinations]\label{thm:linear-combinations}
Suppose that, at the true value \(\bs\theta_0\),
\[
D_0=E_{\bs\theta_0}
\left[\frac{\partial}{\partial\bs\theta^\top}
U(X;\bs\theta_0)\right],
\qquad
\Sigma_0=\operatorname{Var}_{\bs\theta_0}
[U(X;\bs\theta_0)]
\]
exist, \(\Sigma_0\) is positive definite, and
\(W_0D_0\) is nonsingular, where \(W_0=W(\bs\theta_0)\). Under the local
conditions for a consistent selected root,
\[
\sqrt n(\hat{\bs\theta}_W-\bs\theta_0)
\overset{D}{\longrightarrow}\ N_s(\bs0,V_W),
\]
where
\[
V_W=(W_0D_0)^{-1}W_0\Sigma_0W_0^\top(W_0D_0)^{-T}.
\]
\end{theorem}

\begin{proof}
For \(\psi_W=WU\), the terms containing derivatives of \(W\) vanish from
\[
E_{\bs\theta_0}[\partial\psi_W(X;\bs\theta_0)/
\partial\bs\theta^\top],
\]
because
\(E_{\bs\theta_0}[U(X;\bs\theta_0)]=\bs0\). Thus the sensitivity is
\(W_0D_0\), while the variability is
\(W_0\Sigma_0W_0^\top\). The result follows from the standard expansion in
Theorem~\ref{coprinc}.
\end{proof}

\begin{theorem}[Godambe-optimal combination]\label{thm:godambe}
Assume that \(\Sigma_0\) is positive definite and that \(D_0\) has full
column rank. Among matrices \(W_0\in\mathbb R^{s\times q}\) for which
\(W_0D_0\) is nonsingular, \(V_W\) is minimized in the Loewner ordering by
\[
W_{\operatorname{opt}}=D_0^\top\Sigma_0^{-1}.
\]
The minimum covariance and the associated Godambe information are
\[
V_{\operatorname{opt}}
=\left(D_0^\top\Sigma_0^{-1}D_0\right)^{-1},
\qquad
G_0=D_0^\top\Sigma_0^{-1}D_0.
\]
\end{theorem}

\begin{proof}
Let \(L=(W_0D_0)^{-1}W_0\), so that \(LD_0=I_s\), and set
\[
L_0=(D_0^\top\Sigma_0^{-1}D_0)^{-1}
D_0^\top\Sigma_0^{-1}.
\]
Then \(L_0D_0=I_s\), and the cross terms vanish because
\((L-L_0)D_0=0\). Hence
\[
L\Sigma_0L^\top-L_0\Sigma_0L_0^\top
=(L-L_0)\Sigma_0(L-L_0)^\top\succeq0,
\]
which proves both assertions.
\end{proof}

\begin{proposition}[Efficiency bound]\label{prop:fisher-bound}
Let \(S(X;\bs\theta)=\partial\log f(X;\bs\theta)/\partial\bs\theta\)
be the score and let
\(I_0=\operatorname{Var}_{\bs\theta_0}[S(X;\bs\theta_0)]\). Under the
usual differentiability conditions,
\[
G_0\preceq I_0,
\qquad
V_{\operatorname{opt}}\succeq I_0^{-1}.
\]
Equality holds if and only if the score belongs almost surely to the linear
span of the components of \(U(X;\bs\theta_0)\).
\end{proposition}

\begin{proof}
Differentiating \(E_{\bs\theta}[U(X;\bs\theta)]=\bs0\) gives
\(D_0=-\operatorname{Cov}_{\bs\theta_0}(U,S)\). Therefore
\[
I_0-D_0^\top\Sigma_0^{-1}D_0
=\operatorname{Var}_{\bs\theta_0}
\left[S+D_0^\top\Sigma_0^{-1}U\right]\succeq0.
\]
The equality condition is exactly the vanishing of this residual.
\end{proof}

\begin{corollary}[When an explicit estimator is first-order efficient]
Under the conditions of Proposition~\ref{prop:fisher-bound}, suppose there is
a nonsingular matrix \(C_0\) such that
\(S(X;\bs\theta_0)=C_0U(X;\bs\theta_0)\) almost surely. If the corresponding
estimating equations admit an explicit locally identified root, then that AE
estimator has asymptotic covariance \(I_0^{-1}\). Conversely, within the
linear span of \(U\), first-order likelihood efficiency is impossible unless
the score has this representation.
\end{corollary}

\begin{proof}
The score-span condition is the equality condition in
Proposition~\ref{prop:fisher-bound}; the converse follows from the same
condition.
\end{proof}

These optimality statements quantify a trade-off rather than guaranteeing an
explicit optimum. If \(W_{\operatorname{opt}}\) depends nonlinearly on the
unknown parameter, inserting it into the equations can destroy closed-form
solvability. It can instead be used to compare explicit candidates, or with a
preliminary consistent estimator in a two-step procedure.

For practical inference, the sensitivity and variability can be estimated by
\[
\hat D=\frac1n\sum_{i=1}^n
\frac{\partial}{\partial\bs\theta^\top}
U(X_i;\hat{\bs\theta}),
\qquad
\hat\Sigma=\frac1n\sum_{i=1}^n
U(X_i;\hat{\bs\theta})U(X_i;\hat{\bs\theta})^\top.
\]
For the optimal combination, the estimated covariance of
\(\sqrt n(\hat{\bs\theta}-\bs\theta_0)\) is
\[
(\hat D^\top\hat\Sigma^{-1}\hat D)^{-1}.
\]
The influence function for a
fixed regular combination is
\[
\operatorname{IF}(x;\hat{\bs\theta}_W)
=-(W_0D_0)^{-1}W_0U(x;\bs\theta_0),
\]
so bounded components of \(U\) yield bounded influence. Under
misspecification, the selected root instead targets
\(\bs\theta_P\) satisfying \(E_P[\psi_W(X;\bs\theta_P)]=\bs0\), with the
usual sandwich covariance \(A_P^{-1}B_PA_P^{-T}\), where
\[
A_P=E_P\left[\frac{\partial}{\partial\bs\theta^\top}
\psi_W(X;\bs\theta_P)\right],
\qquad
B_P=\operatorname{Var}_P[\psi_W(X;\bs\theta_P)].
\]

\section{Closed-form constructions across model classes}

The first two constructions connect the framework to distributional examples
for which closed-form likelihood-based estimators have previously been studied.
They provide algebraic checks and show how a single auxiliary coordinate can
cover an entire transformed family. The subsequent regression, matrix,
zero-inflated, tail and copula constructions illustrate extensions beyond
those classical settings.

\subsection{Transformed and weighted exponential families}\label{sec:transformed}

Let \(X_1,\ldots,X_n\) be iid with density
\begin{equation}\label{fdpweightedexp}
f(x;\mu,\sigma)=
\frac{(\mu\sigma)^{\mu+1}}{(\sigma+\delta)\Gamma(\mu+1)}
\left[1+\delta T(x)\right]\frac{|T'(x)|}{T(x)}
\exp\{-\mu\sigma T(x)+\mu\log T(x)\},
\end{equation}
where \(x\in D\subseteq(0,\infty)\), \(\mu,\sigma>0\),
\(\delta\in\{0,1\}\), and \(T:D\to(0,\infty)\) is strictly monotone and
twice differentiable \citep{vila2024novel}. The value \(\delta=0\) gives the
non-weighted family and \(\delta=1\) the weighted family. Table
\ref{tab:weighted-exp-cases} records several particular cases; the Gamma and
inverse Gamma rows are retained as members of the general construction, not as
stand-alone examples.

\begin{table}[!ht]
\centering
\caption{Particular cases of the transformed exponential family in
\eqref{fdpweightedexp}, adapted from \citet{vila2024novel}.}
\label{tab:weighted-exp-cases}
\resizebox{\textwidth}{!}{%
\begin{tabular}{llllll}
\toprule
Class & Distribution & \(T(x)\) & \(T'(x)\) & \(T''(x)\) & Parameter mapping\\
\midrule
\multicolumn{6}{l}{\emph{Non-weighted family} \((\delta=0)\)}\\
 & Gamma & \(x\) & \(1\) & \(0\) & \(\mu=\phi,\ \sigma=1/\lambda\)\\
 & Nakagami & \(x^2\) & \(2x\) & \(2\) & \(\mu=m,\ \sigma=1/\Omega\)\\
 & Inverse Gamma & \(x^{-1}\) & \(-x^{-2}\) & \(2x^{-3}\) & \(\mu=\phi,\ \sigma=\lambda/\phi\)\\
 & Inverse Nakagami & \(x^{-2}\) & \(-2x^{-3}\) & \(6x^{-4}\) & \(\mu=m,\ \sigma=1/\Omega\)\\
\midrule
\multicolumn{6}{l}{\emph{Weighted family} \((\delta=1)\)}\\
 & Weighted Lindley & \(x\) & \(1\) & \(0\) & \(\mu=\phi,\ \sigma=\lambda/\phi\)\\
 & Weighted inverse Lindley & \(x^{-1}\) & \(-x^{-2}\) & \(2x^{-3}\) & \(\mu=\phi,\ \sigma=\lambda/\phi\)\\
 & Weighted exponentiated Lindley & \(\log(1+x)\) & \((1+x)^{-1}\) & \(-(1+x)^{-2}\) & \(\mu=\phi,\ \sigma=\lambda/\phi\)\\
 & Weighted log-Lindley & \(\exp(x)-1\) & \(\exp(x)\) & \(\exp(x)\) & \(\mu=\phi,\ \sigma=\lambda/\phi\)\\
 & Weighted Nakagami & \(x^2\) & \(2x\) & \(2\) & \(\mu=m,\ \sigma=1/\Omega\)\\
\bottomrule
\end{tabular}}
\end{table}

Introduce the power-transformed auxiliary density
\begin{equation}\label{gweightedexp}
\begin{aligned}
g(y;\mu,\sigma,p)
&=p\frac{(\mu\sigma)^{\mu+1}}
{(\sigma+\delta)\Gamma(\mu+1)}
\left[1+\delta T(y^p)\right]
\frac{|T'(y^p)|y^{p-1}}{T(y^p)}\\
&\quad{}\times
\exp\{-\mu\sigma T(y^p)+\mu\log T(y^p)\},
\end{aligned}
\end{equation}
where \(p>0\) and \(y^p\in D\). We assume a common support in a
neighbourhood of \(p=1\), or equivalently regular boundary behaviour that
justifies differentiating the normalization. Since
\(g(y;\mu,\sigma,1)=f(y;\mu,\sigma)\), \(p\) is an auxiliary coordinate
fixed at \(p_0=1\). We estimate \((\sigma,\mu)\) using the equation in
\(\sigma\) and the auxiliary equation in \(p\).

For compactness, define
\[
\begin{aligned}
A_1&=\frac1n\sum_i\log X_i,
&A_2&=\frac1n\sum_i\frac{T''(X_i)}{T'(X_i)}X_i\log X_i,\\
A_3&=\frac1n\sum_i\frac{T'(X_i)}{1+T(X_i)}X_i\log X_i,
&A_4&=\frac1n\sum_i\frac{T'(X_i)}{T(X_i)}X_i\log X_i,\\
A_5&=\frac1n\sum_iT'(X_i)X_i\log X_i,
&A_6&=\frac1n\sum_iT(X_i).
\end{aligned}
\]
For \(\delta=0\), direct solution gives
\begin{equation}\label{sigmanonweighted}
\hat\sigma_{\mathrm{AE}}=A_6^{-1},
\qquad
\hat\mu_{\mathrm{AE}}
=\frac{A_6(1+A_1+A_2-A_4)}{A_5-A_6A_4}.
\end{equation}
For \(\delta=1\), put \(Z=1+A_1+A_2+A_3-A_4\). The admissible positive
root is
\begin{equation}\label{sigmaweighted}
\hat\sigma_{\mathrm{AE}}
=\frac{Z(1-A_6)+A_5+
\sqrt{[Z(1-A_6)+A_5]^2+4ZA_6(Z-A_4)}}{2ZA_6},
\end{equation}
and
\begin{equation}\label{muweighted}
\hat\mu_{\mathrm{AE}}
=\frac{Z}{\hat\sigma_{\mathrm{AE}}A_5-A_4}.
\end{equation}
These formulas define the selected branch when the discriminant is
nonnegative and both estimates are positive. At an interior population value,
the corresponding event has probability tending to one whenever the local
moment map is nonsingular.

For example, \(T(x)=x\), \(\delta=1\), \(\mu=\phi\) and
\(\sigma=\lambda/\phi\) give the weighted Lindley density
\citep{kim2021new}
\[
f(x;\lambda,\phi)
=\frac{\lambda^{\phi+1}}{(\lambda+\phi)\Gamma(\phi)}
(1+x)x^{\phi-1}e^{-\lambda x},\qquad x>0.
\]
Here
\[
\begin{aligned}
Z&=1+\frac1n\sum_i\frac{X_i\log X_i}{1+X_i},
&A_4&=\frac1n\sum_i\log X_i,\\
A_5&=\frac1n\sum_iX_i\log X_i,
&A_6&=\bar X.
\end{aligned}
\]
and \(\hat\phi_{\mathrm{AE}}=Z/(\hat\sigma_{\mathrm{AE}}A_5-A_4)\),
\(\hat\lambda_{\mathrm{AE}}=\hat\sigma_{\mathrm{AE}}
\hat\phi_{\mathrm{AE}}\), with \(\hat\sigma_{\mathrm{AE}}\) from
\eqref{sigmaweighted}.

\subsection{Beta distribution}

For \(X\sim\operatorname{Beta}(\alpha,\beta)\), the likelihood equations
involve digamma functions. Consider instead the generalized Beta density
\[
g(x;\alpha,\beta,a,c)
=\frac{(x-a)^{\alpha-1}(c-x)^{\beta-1}}
{(c-a)^{\alpha+\beta-1}\operatorname{B}(\alpha,\beta)},
\qquad a<x<c,
\]
and evaluate the two auxiliary endpoint derivatives at \((a,c)=(0,1)\).
For \(\alpha>1\) and \(\beta>1\),
\[
E\left(\frac1X\right)=\frac{\alpha+\beta-1}{\alpha-1},
\qquad
E\left(\frac1{1-X}\right)=\frac{\alpha+\beta-1}{\beta-1},
\]
so both centered auxiliary equations have zero expectation. Put
\[
S_1=\sum_iX_i^{-1},\quad
S_2=\sum_i(1-X_i)^{-1},\quad
R_1=S_1-n,\quad R_2=S_2-n.
\]
If the observations are not all equal, Cauchy--Schwarz gives
\(R_1R_2>n^2\), and the unique positive solution is
\begin{equation}\label{eq:beta-ae}
\hat\alpha_{\mathrm{AE}}
=\frac{S_1}{R_1-n^2/R_2},
\qquad
\hat\beta_{\mathrm{AE}}
=\frac{S_2}{R_2-n^2/R_1}.
\end{equation}

\begin{proposition}[Joint limit for the Beta estimator]\label{prop:beta-limit}
If \(\alpha>2\) and \(\beta>2\), the estimator in
\eqref{eq:beta-ae} is strongly consistent and
\[
\sqrt n
\begin{pmatrix}
\hat\alpha_{\mathrm{AE}}-\alpha\\
\hat\beta_{\mathrm{AE}}-\beta
\end{pmatrix}
\overset{D}{\longrightarrow}
N_2\left(\bs0,
\begin{pmatrix}
Q(\alpha,\beta)&R(\alpha,\beta)\\
R(\alpha,\beta)&Q(\beta,\alpha)
\end{pmatrix}\right),
\]
where
\[
Q(y,z)=
\frac{y(y-1)^2\{2yz-3y+2z^2-8z+6\}}
{(y-2)(z-2)(y+z-1)}
\]
and
\[
\begin{aligned}
R(y,z)
&=\frac{(y-1)(z-1)}{(y-2)(z-2)(y+z-1)}\\
&\quad\times
\left\{2yz(y+z)-2(y^2+z^2)-9yz+6(y+z)-4\right\}.
\end{aligned}
\]
\end{proposition}

\begin{proof}
The estimator is a continuous function of
\(\bar Z_n=n^{-1}\sum_i Z_i\), where
\(Z_i=(X_i^{-1},(1-X_i)^{-1})^\top\), in a neighbourhood of its population
value. The strong law gives consistency. Its covariance matrix is
\[
\Sigma_Z=(\alpha+\beta-1)
\begin{pmatrix}
\dfrac{\beta}{(\alpha-2)(\alpha-1)^2}&
-\dfrac1{(\alpha-1)(\beta-1)}\\[6pt]
-\dfrac1{(\alpha-1)(\beta-1)}&
\dfrac{\alpha}{(\beta-2)(\beta-1)^2}
\end{pmatrix}.
\]
The Jacobian of the inverse map at the population moments is
\[
-\frac1{\alpha+\beta-1}
\begin{pmatrix}
\alpha(\alpha-1)^2&(\alpha-1)(\beta-1)^2\\
(\alpha-1)^2(\beta-1)&\beta(\beta-1)^2
\end{pmatrix}.
\]
The multivariate central limit theorem and delta method give the stated matrix
after multiplication. In particular, the off-diagonal covariance between the
two reciprocal moments is essential for both marginal variances.
\end{proof}

\subsection{Weibull accelerated failure time regression}\label{sec:aft}

We illustrate the framework with a Weibull accelerated failure time (AFT)
regression model \citep{kalbfleisch2002statistical}. This example is useful
because the ordinary likelihood
equations for the regression coefficients and shape parameter are coupled
nonlinearly, while the proposed auxiliary construction gives an explicit
estimator that can be reused inside larger algorithms.

Let \(T_1,\ldots,T_n\) be independent positive random variables and let
\(\bs x_i\in\mathbb R^p\) be fixed covariates, including an intercept. Assume
\[
T_i\mid \bs x_i \sim \operatorname{Weibull}(k,\lambda_i),
\qquad
k>0,
\qquad
\lambda_i=\exp(\bs x_i^\top\bs\beta),
\]
with density
\[
f(t_i;\bs\beta,k)
=
\frac{k}{\lambda_i}
\left(\frac{t_i}{\lambda_i}\right)^{k-1}
\exp\left\{
-\left(\frac{t_i}{\lambda_i}\right)^k
\right\},
\qquad t_i>0.
\]
Let \(Y_i=\log T_i\). Then
\[
Y_i=\bs x_i^\top\bs\beta+\frac1k\varepsilon_i,
\qquad
\varepsilon_i=\log E_i,\quad E_i\sim\operatorname{Exp}(1).
\]
Hence \(E(\varepsilon_i)=-\gamma\) and
\(\operatorname{Var}(\varepsilon_i)=\pi^2/6\), where \(\gamma\) is the
Euler--Mascheroni constant. Therefore,
\[
E(Y_i\mid \bs x_i)
=
\bs x_i^\top\bs\beta-\frac{\gamma}{k},
\qquad
\operatorname{Var}(Y_i\mid \bs x_i)
=
\frac{\pi^2}{6k^2}.
\]

Let \(X_d=(\bs x_1^\top,\ldots,\bs x_n^\top)^\top\),
\(\bs Y=(Y_1,\ldots,Y_n)^\top\), and assume that \(X_d^\top X_d\) is
nonsingular. Define
\[
\tilde{\bs\beta}=(X_d^\top X_d)^{-1}X_d^\top\bs Y,
\qquad
RSS=\sum_{i=1}^n
\left(Y_i-\bs x_i^\top\tilde{\bs\beta}\right)^2.
\]
Since the intercept absorbs the term \(-\gamma/k\), a closed-form estimator of
\(k\) is
\[
\hat k_{\mathrm{AE}}
=
\frac{\pi}{\sqrt{6\,RSS/n}}.
\]
A degrees-of-freedom corrected version may replace \(RSS/n\) by \(RSS/(n-p)\).
With \(\bs e_1=(1,0,\ldots,0)^\top\), the intercept-corrected estimator is
\[
\hat{\bs\beta}_{\mathrm{AE}}
=
\tilde{\bs\beta}
+
\frac{\gamma}{\hat k_{\mathrm{AE}}}\bs e_1.
\]

To embed this estimator in the auxiliary-equation construction,
define
\[
U_1(t_i,\bs x_i;\bs\beta,k)
=
\bs x_i
\left(
\log t_i-\bs x_i^\top\bs\beta+\frac{\gamma}{k}
\right)
\]
and
\[
U_2(t_i,\bs x_i;\bs\beta,k)
=
\left(
\log t_i-\bs x_i^\top\bs\beta+\frac{\gamma}{k}
\right)^2
-
\frac{\pi^2}{6k^2}.
\]
These functions have conditional mean zero under the Weibull AFT model. The
estimating-function generator in Section~\ref{sec:theory} therefore gives
\[
g_i(t_i;\bs\beta,k,\bs a,b)
=
f(t_i;\bs\beta,k)
\exp\left\{
\bs a^\top U_1(t_i,\bs x_i;\bs\beta,k)
+
bU_2(t_i,\bs x_i;\bs\beta,k)
\right\},
\]
where \((\bs a,b)\) are auxiliary coordinates fixed at \((\bs0,0)\) after
differentiation. The resulting auxiliary estimating equations are
\[
\sum_{i=1}^n
\bs x_i
\left(
\log T_i-\bs x_i^\top\bs\beta+\frac{\gamma}{k}
\right)
=0
\]
and
\[
\sum_{i=1}^n
\left[
\left(
\log T_i-\bs x_i^\top\bs\beta+\frac{\gamma}{k}
\right)^2
-
\frac{\pi^2}{6k^2}
\right]
=0,
\]
whose solution is exactly the closed-form estimator above. In contrast, the
ordinary likelihood equations involve nonlinear terms of the form
\((T_i/\exp\{\bs x_i^\top\bs\beta\})^k\) and do not admit a closed-form
solution in general.

The next result gives an explicit joint limit, rather than appealing only to
the generic independent non-identically distributed proposition. Let
\(\zeta(3)=\sum_{j\geq1}j^{-3}\).

\begin{proposition}[Joint limit for Weibull AFT]\label{prop:aft-limit}
Suppose that \(p\) is fixed, the first column of \(X_d\) is one,
\(n^{-1}X_d^\top X_d\to Q\) for a positive definite matrix \(Q\), and
\(\max_i\|\bs x_i\|/\sqrt n\to0\). Then
\[
\sqrt n
\begin{pmatrix}
\hat{\bs\beta}_{\mathrm{AE}}-\bs\beta\\
\hat k_{\mathrm{AE}}-k
\end{pmatrix}
\overset{D}{\longrightarrow}
N_{p+1}\left(\bs0,
\begin{pmatrix}
V_{\beta\beta}&V_{\beta k}\\
V_{\beta k}^\top&V_{kk}
\end{pmatrix}\right),
\]
where
\[
\begin{aligned}
V_{\beta\beta}
&=\frac1{k^2}\left[
\frac{\pi^2}{6}Q^{-1}
+\left\{\frac{11\gamma^2}{10}
-\frac{12\gamma\zeta(3)}{\pi^2}\right\}
\bs e_1\bs e_1^\top\right],\\
V_{\beta k}
&=\left\{\frac{6\zeta(3)}{\pi^2}
-\frac{11\gamma}{10}\right\}\bs e_1,
\qquad
V_{kk}=\frac{11}{10}k^2.
\end{aligned}
\]
Replacing \(RSS/n\) by \(RSS/(n-p)\) leaves this first-order limit unchanged.
\end{proposition}

\begin{proof}
Write \(\eta_i=\varepsilon_i+\gamma\). Then
\[
E(\eta_i)=0,\quad E(\eta_i^2)=\frac{\pi^2}{6},\quad
E(\eta_i^3)=-2\zeta(3),\quad
E(\eta_i^4)=\frac{3\pi^4}{20}.
\]
The OLS estimator targets
\(\bs\beta-\gamma\bs e_1/k\). Since \(p\) is fixed, replacing the true
centered errors by OLS residuals changes their mean square by
\(O_p(n^{-1})\). A joint Lindeberg--Feller central limit theorem applied to
\(\bs x_i\eta_i\) and
\(\eta_i^2-\pi^2/6\), followed by the delta method for
\(k=\{(\pi^2/6)/(\pi^2/(6k^2))\}^{1/2}\), gives
\[
\operatorname{avar}\{\sqrt n(\hat k_{\mathrm{AE}}-k)\}
=\frac{k^2}{4(\pi^2/6)^2}
\operatorname{Var}(\eta_i^2)=\frac{11k^2}{10}.
\]
Using \(Q^{-1}\lim_n n^{-1}\sum_i\bs x_i=\bs e_1\), the remaining
blocks follow from the third central moment and the intercept correction
\(\gamma/\hat k_{\mathrm{AE}}\).
\end{proof}

\subsection{Wishart covariance model}
Let \(X_1,\ldots,X_n\) be iid random matrices following a Wishart distribution,
\(X_i\sim W_d(\nu,\Sigma)\), with \(\nu>d-1\) and
\(\Sigma\in\mathcal S_d^{++}\) \citep{muirhead1982aspects}. Its density is
\[
f(X;\nu,\Sigma)
=
\frac{
|X|^{(\nu-d-1)/2}
\exp\left\{-\frac12\operatorname{tr}(\Sigma^{-1}X)\right\}
}{
2^{\nu d/2}
|\Sigma|^{\nu/2}
\Gamma_d(\nu/2)
},
\qquad X\in\mathcal S_d^{++},
\]
where \(\Gamma_d(\cdot)\) is the multivariate gamma function. We use the mean
parametrization
\[
M=E(X)=\nu\Sigma,
\qquad
\Sigma=\frac{M}{\nu}.
\]
Under this parametrization,
\[
f(X;\nu,M)
=
\frac{
|X|^{(\nu-d-1)/2}
\exp\left\{-\frac{\nu}{2}\operatorname{tr}(M^{-1}X)\right\}
}{
2^{\nu d/2}
|M/\nu|^{\nu/2}
\Gamma_d(\nu/2)
}.
\]

For fixed \(\nu\), the usual maximum likelihood estimator of \(\Sigma\) is
\(\bar X/\nu\), where
\[
\bar X=\frac1n\sum_{i=1}^n X_i.
\]
However, the likelihood equation for \(\nu\) involves the multivariate
digamma function and does not admit a closed-form solution. We now construct
a closed-form auxiliary-equation estimator.

The key identities are
\[
E_{\nu,M}(X)=M
\]
and
\[
E_{\nu,M}\left[X\log |X|\right]
-
E_{\nu,M}(X)E_{\nu,M}\left[\log |X|\right]
=
\frac{2}{\nu}M.
\]
Consequently,
\[
E_{\nu,M}
\left[
\operatorname{tr}\left\{
M^{-1}(X-M)
\right\}
\log |X|
\right]
=
\frac{2d}{\nu}.
\]

Define
\[
U_1(X;M)=\operatorname{vech}(X-M)
\]
and
\[
U_2(X;\nu,M)
=
\operatorname{tr}\left\{M^{-1}(X-M)\right\}\log |X|
-
\frac{2d}{\nu}.
\]
Then
\[
E_{\nu,M}[U_1(X;M)]=0,
\qquad
E_{\nu,M}[U_2(X;\nu,M)]=0.
\]
By the estimating-function generator, define
\[
\begin{aligned}
g(X;\nu,M,\bs a,b)
&=
f(X;\nu,M)
\exp\left\{
\bs a^\top \operatorname{vech}(X-M)\right.\\
&\qquad\left.
+
b\left[
\operatorname{tr}\left\{M^{-1}(X-M)\right\}\log |X|
-\frac{2d}{\nu}
\right]
\right\},
\end{aligned}
\]
where \(\bs a\in\mathbb R^{d(d+1)/2}\), \(b\in\mathbb R\), and
\((\bs a_0,b_0)=(\bs0,0)\). The auxiliary estimating equations over
\((\bs a,b)\), evaluated at \((\bs0,0)\), are
\[
\frac1n\sum_{i=1}^n\operatorname{vech}(X_i-M)=0
\]
and
\[
\frac1n\sum_{i=1}^n
\operatorname{tr}\left\{M^{-1}(X_i-M)\right\}
\log |X_i|
-
\frac{2d}{\nu}
=0.
\]

The first equation gives
\[
\hat M_{\mathrm{AE}}=\bar X.
\]
Substituting \(\hat M_{\mathrm{AE}}\) into the second equation gives
\[
\hat\nu_{\mathrm{AE}}
=
\frac{
2d
}{
\frac1n\sum_{i=1}^n
\operatorname{tr}
\left\{
\bar X^{-1}(X_i-\bar X)
\right\}
\log |X_i|
}.
\]
Equivalently,
\[
\hat\nu_{\mathrm{AE}}
=
\frac{
2d
}{
\operatorname{tr}
\left[
\bar X^{-1}
\left(
\frac1n\sum_{i=1}^n X_i\log |X_i|
-
\bar X\frac1n\sum_{i=1}^n\log |X_i|
\right)
\right]
}.
\]
Finally,
\[
\hat\Sigma_{\mathrm{AE}}
=
\frac{\hat M_{\mathrm{AE}}}{\hat\nu_{\mathrm{AE}}}
=
\frac{\bar X}{\hat\nu_{\mathrm{AE}}}.
\]

\begin{proposition}[Wishart large-sample limit]\label{prop:wishart-limit}
Let
\[
\tau(X)=
\begin{pmatrix}
\operatorname{vech}(X)\\
\operatorname{vech}(X\log|X|)\\
\log|X|
\end{pmatrix},
\qquad
\Omega_\tau=\operatorname{Var}_{\nu,M}[\tau(X)].
\]
For \(m,c\in\mathbb R^{d(d+1)/2}\) and \(\ell\in\mathbb R\), write
\(M_m=\operatorname{vech}^{-1}(m)\),
\(C_c=\operatorname{vech}^{-1}(c)\), and define
\[
\Delta(m,c,\ell)=\operatorname{tr}
\{M_m^{-1}(C_c-M_m\ell)\},
\qquad
\Phi(m,c,\ell)=\begin{pmatrix}m\\2d/\Delta(m,c,\ell)\end{pmatrix}.
\]
Then \((\operatorname{vech}(\hat M_{\mathrm{AE}})^\top,
\hat\nu_{\mathrm{AE}})^\top=\Phi(\bar\tau_n)\). For every
\(\nu>d-1\) and \(M\in\mathcal S_d^{++}\),
\[
\sqrt n\left[
\Phi(\bar\tau_n)-
\begin{pmatrix}\operatorname{vech}(M)\\\nu\end{pmatrix}
\right]
\overset{D}{\longrightarrow}
N\left(\bs0,
\dot\Phi\,\Omega_\tau\dot\Phi^\top\right),
\]
where \(\dot\Phi\) is evaluated at \(E_{\nu,M}[\tau(X)]\). The limit for
\(\hat\Sigma_{\mathrm{AE}}=\hat M_{\mathrm{AE}}/\hat\nu_{\mathrm{AE}}\)
follows by one further delta-method transformation.
\end{proposition}

\begin{proof}
Wishart variables and their log determinants have finite moments of every
order needed here. The multivariate law of large numbers and central limit
theorem therefore apply to \(\bar\tau_n\). At the population value,
\(\Delta\{E\tau(X)\}=2d/\nu>0\), and the mean matrix is positive definite, so
\(\Phi\) is continuously differentiable in a neighbourhood of that value.
The result follows from the delta method.
\end{proof}

The denominator converges in probability to \(2d/\nu>0\), so the estimator is
well defined and consistent with probability tending to one. For finite-sample
stabilization, one may project onto a specified closed interval
\([d-1+\epsilon,\nu_{\max}]\), where \(\epsilon>0\) and
\(\nu_{\max}<\infty\) are chosen so that the true value is interior. Proposition
\ref{prop:projection} then applies. This example shows that the framework is
not restricted to scalar observations: matrix-valued likelihoods can also be
paired with explicit auxiliary equations.

\subsection{Zero-inflated Poisson model for count data}
Let \(X_1,\ldots,X_n\) be iid random variables following a zero-inflated
Poisson distribution, denoted by \(X\sim \operatorname{ZIP}(\pi,\lambda)\),
with \(\lambda>0\) and \(0\leq\pi<1\) \citep{lambert1992zero}. Its
probability mass function is
\[
p(x;\pi,\lambda)=
\begin{cases}
\pi+(1-\pi)e^{-\lambda}, & x=0,\\
(1-\pi)e^{-\lambda}\lambda^x/x!, & x=1,2,\ldots.
\end{cases}
\]
The first two factorial moments are
\[
E(X)=(1-\pi)\lambda,
\qquad
E[X(X-1)]=(1-\pi)\lambda^2.
\]
Thus define
\[
U_1(x;\pi,\lambda)=x-(1-\pi)\lambda,
\qquad
U_2(x;\pi,\lambda)=x(x-1)-(1-\pi)\lambda^2.
\]
Since
\[
E_{\pi,\lambda}[U_1(X;\pi,\lambda)]
=
E_{\pi,\lambda}[U_2(X;\pi,\lambda)]
=
0,
\]
the estimating-function generator gives
\[
g(x;\pi,\lambda,a_1,a_2)
=
p(x;\pi,\lambda)
\exp\{a_1U_1(x;\pi,\lambda)+a_2U_2(x;\pi,\lambda)\}.
\]
The auxiliary estimating equations over \((a_1,a_2)\), evaluated at
\((a_1,a_2)=(0,0)\), are
\[
\bar X=(1-\pi)\lambda,
\qquad
M_2=(1-\pi)\lambda^2,
\qquad
M_2=\frac1n\sum_{i=1}^nX_i(X_i-1).
\]
Solving this system gives
\[
\hat\lambda_{\mathrm{AE}}
=
\frac{M_2}{\bar X},
\qquad
\hat\pi_{\mathrm{AE}}
=
1-\frac{\bar X^2}{M_2},
\]
provided that \(\bar X>0\) and \(M_2>0\).

\begin{proposition}[Limit for the ZIP estimator]\label{prop:zip-limit}
If \(0<\pi<1\) and \(\lambda>0\), then the unrestricted estimator is
well defined with probability tending to one, is consistent, and
\[
\sqrt n
\begin{pmatrix}
\hat\pi_{\mathrm{AE}}-\pi\\
\hat\lambda_{\mathrm{AE}}-\lambda
\end{pmatrix}
\overset{D}{\longrightarrow}
N_2\left(\bs0,
\begin{pmatrix}
(1-\pi)\left(\pi+2/\lambda^2\right)&2/\lambda\\
2/\lambda&(\lambda+2)/(1-\pi)
\end{pmatrix}\right).
\]
\end{proposition}

\begin{proof}
For \(Y_1=X\), \(Y_2=X(X-1)\), and \(q=1-\pi\), the factorial-moment
identities give
\[
\operatorname{Var}\begin{pmatrix}Y_1\\Y_2\end{pmatrix}
=q\begin{pmatrix}
\lambda+\pi\lambda^2&2\lambda^2+\pi\lambda^3\\
2\lambda^2+\pi\lambda^3&2\lambda^2+4\lambda^3+\pi\lambda^4
\end{pmatrix}.
\]
The inverse moment map is
\((m_1,m_2)\mapsto(1-m_1^2/m_2,m_2/m_1)\). Applying
Proposition~\ref{prop:moment-limit} and multiplying by its Jacobian gives the
displayed covariance matrix.
\end{proof}

If the unrestricted estimate falls
outside the parameter space, the pair
\((\hat\pi_{\mathrm{AE}},\hat\lambda_{\mathrm{AE}})\) may be projected onto
\([0,1-\epsilon]\times[\epsilon,\lambda_{\max}]\), with fixed
\(\epsilon>0\). For an interior true value, this projection is inactive with
probability tending to one. The ordinary MLE for the ZIP model is obtained from
nonlinear equations involving the zero-count probability
\(\pi+(1-\pi)e^{-\lambda}\), and numerical optimization is typically
required.

\subsection{Generalized Pareto distribution}
Let \(X_1,\ldots,X_n\) be iid random variables following a generalized Pareto
distribution, \(X\sim\operatorname{GPD}(\sigma,\xi)\), with \(\sigma>0\). For
\(\xi\neq0\), its density is \citep{embrechts1997modelling}
\[
f(x;\sigma,\xi)
=
\frac1\sigma
\left(1+\frac{\xi x}{\sigma}\right)^{-1/\xi-1},
\]
where \(x>0\) if \(\xi\geq0\), and \(0<x<-\sigma/\xi\) if \(\xi<0\). The case
\(\xi=0\) is obtained as the exponential limit. If \(\xi<1/2\), then
\[
E(X)=\frac{\sigma}{1-\xi},
\qquad
\operatorname{Var}(X)=
\frac{\sigma^2}{(1-\xi)^2(1-2\xi)}.
\]
Define
\[
U_1(x;\sigma,\xi)
=
x-\frac{\sigma}{1-\xi}
\]
and
\[
U_2(x;\sigma,\xi)
=
\left(x-\frac{\sigma}{1-\xi}\right)^2
-
\frac{\sigma^2}{(1-\xi)^2(1-2\xi)}.
\]
Then
\[
E_{\sigma,\xi}[U_1(X;\sigma,\xi)]
=
E_{\sigma,\xi}[U_2(X;\sigma,\xi)]
=
0,
\]
and we may take
\[
g(x;\sigma,\xi,a_1,a_2)
=
f(x;\sigma,\xi)
\exp\{a_1U_1(x;\sigma,\xi)+a_2U_2(x;\sigma,\xi)\}.
\]
The auxiliary estimating equations over \((a_1,a_2)\), evaluated at
\((a_1,a_2)=(0,0)\), are
\[
\bar X=\frac{\sigma}{1-\xi},
\qquad
S_X^2=
\frac{\sigma^2}{(1-\xi)^2(1-2\xi)},
\qquad
S_X^2=\frac1n\sum_{i=1}^n(X_i-\bar X)^2.
\]
Solving this system gives
\[
\hat\xi_{\mathrm{AE}}
=
\frac12\left(1-\frac{\bar X^2}{S_X^2}\right),
\qquad
\hat\sigma_{\mathrm{AE}}
=
\bar X(1-\hat\xi_{\mathrm{AE}}).
\]

\begin{proposition}[Limit for the GPD moment estimator]\label{prop:gpd-limit}
Suppose \(\xi<1/4\), and let \(\mu=\sigma/(1-\xi)\) and
\(v=\sigma^2/\{(1-\xi)^2(1-2\xi)\}\). Define
\[
\mu_3=\frac{2\sigma^3(1+\xi)}
{(1-\xi)^3(1-2\xi)(1-3\xi)},
\quad
\mu_4=\frac{3\sigma^4(2\xi^2+\xi+3)}
{(1-\xi)^4(1-2\xi)(1-3\xi)(1-4\xi)}.
\]
Then
\[
\sqrt n
\begin{pmatrix}
\hat\sigma_{\mathrm{AE}}-\sigma\\
\hat\xi_{\mathrm{AE}}-\xi
\end{pmatrix}
\overset{D}{\longrightarrow}
N_2(\bs0,D\Omega D^\top),
\]
where
\[
\Omega=\begin{pmatrix}v&\mu_3\\\mu_3&\mu_4-v^2\end{pmatrix},
\qquad
D=\begin{pmatrix}
\dfrac12+\dfrac{3\mu^2}{2v}&-\dfrac{\mu^3}{2v^2}\\[6pt]
-\dfrac{\mu}{v}&\dfrac{\mu^2}{2v^2}
\end{pmatrix}.
\]
\end{proposition}

\begin{proof}
The vector consisting of the sample mean and the sample variance has
asymptotic covariance \(\Omega\). The inverse moment map is
\[
(m,v_*)\longmapsto
\left(\frac m2\left\{1+\frac{m^2}{v_*}\right\},
\frac12\left\{1-\frac{m^2}{v_*}\right\}\right),
\]
whose Jacobian at \((\mu,v)\) is \(D\). The result follows from the delta
method. The fourth-moment condition is exactly \(\xi<1/4\).
\end{proof}

The estimator is well defined when \(S_X^2>0\) and the resulting value lies in
the desired admissible region, including the fitted-support restriction when
\(\hat\xi_{\mathrm{AE}}<0\). Consistency of the moment inversion requires
\(\xi<1/2\). Because the second estimating function is quadratic, the
ordinary root-\(n\) normal limit from Theorem~\ref{coprinc} requires a finite
fourth moment, hence \(\xi<1/4\). A constrained version should be defined on a
closed subset of the admissible region that also respects the observed support.
The ordinary MLE generally requires numerical solution of nonlinear
likelihood equations, especially for the tail index \(\xi\).

\subsection{Farlie--Gumbel--Morgenstern copula}
Let \(\{(U_i,V_i)\}_{i=1}^n\) be iid from the Farlie--Gumbel--Morgenstern
copula. Following \citet{nelsen2006introduction}, write its density as
\[
c_\theta(u,v)
=
1+\theta(1-2u)(1-2v),
\qquad
0<u,v<1,\quad -1\leq\theta\leq1.
\]
Put
\[
A(u,v)=(1-2u)(1-2v).
\]
Since
\[
E_\theta[A(U,V)]=\frac{\theta}{9},
\]
define the estimating function
\[
H(u,v;\theta)
=
A(u,v)-\frac{\theta}{9}.
\]
Then \(E_\theta[H(U,V;\theta)]=0\), and the estimating-function generator gives
\[
g(u,v;\theta,a)
=
c_\theta(u,v)
\exp\left\{
a\left[A(u,v)-\frac{\theta}{9}\right]
\right\}.
\]
The auxiliary likelihood equation over \(a\), evaluated at \(a=0\), is
\[
\frac1n\sum_{i=1}^n(1-2U_i)(1-2V_i)-\frac{\theta}{9}=0.
\]
Therefore,
\[
\hat\theta_{\mathrm{AE}}
=
9\cdot\frac1n\sum_{i=1}^n(1-2U_i)(1-2V_i).
\]
Since \(E_\theta[A(U,V)^2]=1/9\),
\[
\sqrt n(\hat\theta_{\mathrm{AE}}-\theta)
\overset{D}{\longrightarrow}N(0,9-\theta^2).
\]
If necessary, a constrained estimator may be obtained by projecting
\(\hat\theta_{\mathrm{AE}}\) onto \([-1,1]\). For data with unknown margins,
\(U_i\) and \(V_i\) may be replaced by pseudo-observations based on scaled
ranks. The ordinary MLE solves
\[
\sum_{i=1}^n
\frac{(1-2U_i)(1-2V_i)}
{1+\theta(1-2U_i)(1-2V_i)}
=0,
\]
which does not admit a closed-form solution in general.

\subsection{Identification and regularity summary}

Table~\ref{tab:regularity-summary} separates three issues that are sometimes
conflated for explicit estimators: algebraic existence on a finite sample,
identification of the population branch, and the moment conditions needed for
a Gaussian root-\(n\) limit. Closed form alone guarantees none of them.

\begin{table}[!ht]
\centering
\caption{Selected equations, admissibility and first-order conditions for the
closed-form constructions.}
\label{tab:regularity-summary}
\small
\setlength{\tabcolsep}{3pt}
\begin{tabular}{@{}>{\raggedright\arraybackslash}p{0.16\textwidth}>{\raggedright\arraybackslash}p{0.22\textwidth}>{\raggedright\arraybackslash}p{0.28\textwidth}>{\raggedright\arraybackslash}p{0.23\textwidth}@{}}
\toprule
Model & Empirical quantities & Identification and admissibility & Root-\(n\) condition\\
\midrule
Transformed family
& \(A_1,\ldots,A_6\)
& Positive selected root, nonzero denominators and a nonsingular local map in
\((\mu,\sigma)\)
& Finite second moments of the two generated estimating functions\\[3pt]
Beta
& \(\overline{X^{-1}}\), \(\overline{(1-X)^{-1}}\)
& Nonconstant sample; local branch is positive for \(\alpha,\beta>1\)
& \(\alpha>2\) and \(\beta>2\)\\[3pt]
Weibull AFT
& OLS coefficients and residual mean square of \(\log T\)
& Full column rank of \(X_d\), intercept included, \(RSS>0\)
& \(n^{-1}X_d^\top X_d\to Q>0\) and negligible maximum leverage\\[3pt]
Wishart
& \(\bar X\), \(\overline{X\log|X|}\), \(\overline{\log|X|}\)
& \(\bar X\in\mathcal S_d^{++}\), positive denominator and \(\nu>d-1\)
& Holds for interior \(\nu>d-1\), \(M\in\mathcal S_d^{++}\)\\[3pt]
ZIP
& First two factorial moments
& \(\bar X>0\), \(M_2>0\), with the population value in
\(0<\pi<1,\lambda>0\)
& All required moments are finite\\[3pt]
GPD
& Sample mean and variance
& \(S_X^2>0\), fitted support respected; consistency for \(\xi<1/2\)
& Finite fourth moment, equivalently \(\xi<1/4\)\\[3pt]
FGM copula
& \(\overline{(1-2U)(1-2V)}\)
& Interior branch \(-1<\theta<1\); projection available on \([-1,1]\)
& All required moments are bounded\\
\bottomrule
\end{tabular}
\end{table}

For the transformed and bounded-parameter examples, the finite-sample event
can fail even though its probability tends to zero at an interior true value.
The estimator must therefore include a branch or stabilization rule in its
definition. For the GPD, the distinction is more substantive: the moment
estimator remains consistent on \(\xi<1/2\), but its usual sandwich covariance
is not valid throughout that region because the quadratic estimating function
has infinite variance when \(1/4\leq\xi<1/2\).

\section{Simulation summary}

We report Monte Carlo experiments for the Beta benchmark and Weibull AFT
regression. For a parameter vector \(\bs\theta\), accuracy is summarized by
\[
\|\operatorname{Bias}(\hat{\bs\theta})\|_2
=\left\|B^{-1}\sum_{b=1}^B\hat{\bs\theta}^{(b)}-\bs\theta\right\|_2
\]
and
\[
\operatorname{RMSE}(\hat{\bs\theta})
=\left[B^{-1}\sum_{b=1}^B
\|\hat{\bs\theta}^{(b)}-\bs\theta\|_2^2\right]^{1/2}.
\]
Computational performance is measured by
\[
\text{Speedup}
=
\frac{\text{MLE computing time}}
{\text{AE computing time}}.
\]
Random number generation and sample summaries common to both estimators are
excluded from the timing comparison. Within each scenario, the two estimators
were implemented in the same computing environment and applied to identical
simulated samples, so the reported timing ratios should be interpreted as
relative algorithmic comparisons rather than hardware-independent constants.

\subsection{Beta distribution}

We first take \((\alpha,\beta)=(3,2.5)\), use \(B=20{,}000\) Monte Carlo
replications, and let \(n=1000,2000,\ldots,10000\). The MLE solves the usual
digamma equations, whereas the AE estimator is given by
\eqref{eq:beta-ae}.

\begin{figure}[!ht]
\centering
\includegraphics[width=\textwidth]{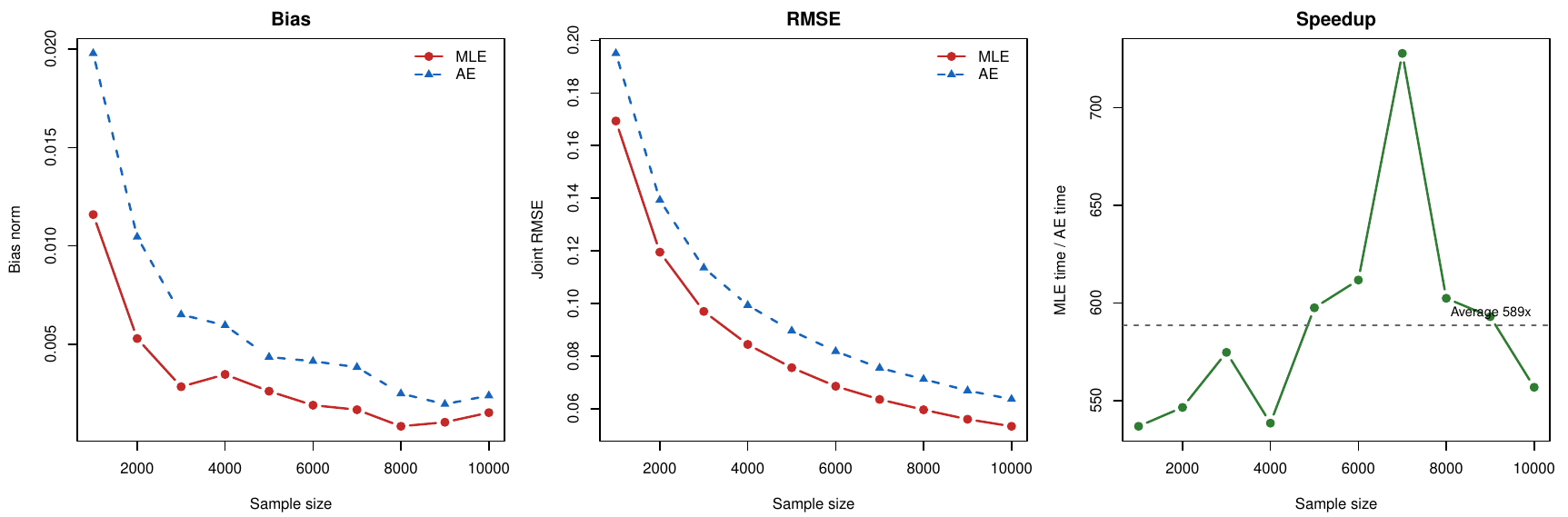}
\caption{Monte Carlo comparison for the Beta parameters
\((\alpha,\beta)\). Left: Euclidean norm of the bias vector. Center: joint
RMSE. Right: computational speedup, defined as MLE time divided by AE
estimator time.}
\label{fig:beta-bias-rmse-speedup}
\end{figure}

Figure~\ref{fig:beta-bias-rmse-speedup} shows the expected efficiency trade-off.
The AE estimator is consistent and its error decreases with \(n\), but its
finite-sample bias and RMSE are larger than those of the MLE. The average joint
RMSE ratio is approximately \(1.18\), while the average computational speedup
is approximately \(589\). Thus the explicit estimator sacrifices moderate
finite-sample accuracy for a substantial reduction in computing time.

\subsection{Weibull AFT regression model}

Finally, we consider the Weibull accelerated failure time model in
Section~\ref{sec:aft}.
The data are generated from
\[
\log T_i
=
\bs x_i^\top\bs\beta+\frac{1}{k}\varepsilon_i,
\qquad
\varepsilon_i=\log E_i,\quad E_i\sim\operatorname{Exp}(1),
\]
with \(k=1.7\). We set
\(\bs\beta=(0.5,-0.6,0.4)^\top\), and the covariate vector is
\(\bs x_i=(1,Z_{i1},Z_{i2})^\top\), where \(Z_{i1}\sim N(0,1)\) and
\(Z_{i2}\sim U(-1,1)\). We use \(B=300\) Monte Carlo replications and sample
sizes \(n=1000,2000,\ldots,10000\). Accuracy is summarized jointly for
\(\bs\theta=(k,\bs\beta^\top)^\top\), using the same bias norm and joint RMSE
as in the Beta simulation.

The MLE is computed by numerical maximization of the Weibull likelihood. The
AE estimator follows from the closed-form log-Weibull construction in
Section~\ref{sec:aft} and only requires an ordinary least-squares fit to
\(\log T_i\),
followed by the residual-variance correction for \(k\).

\begin{figure}[!ht]
\centering
\includegraphics[width=\textwidth]{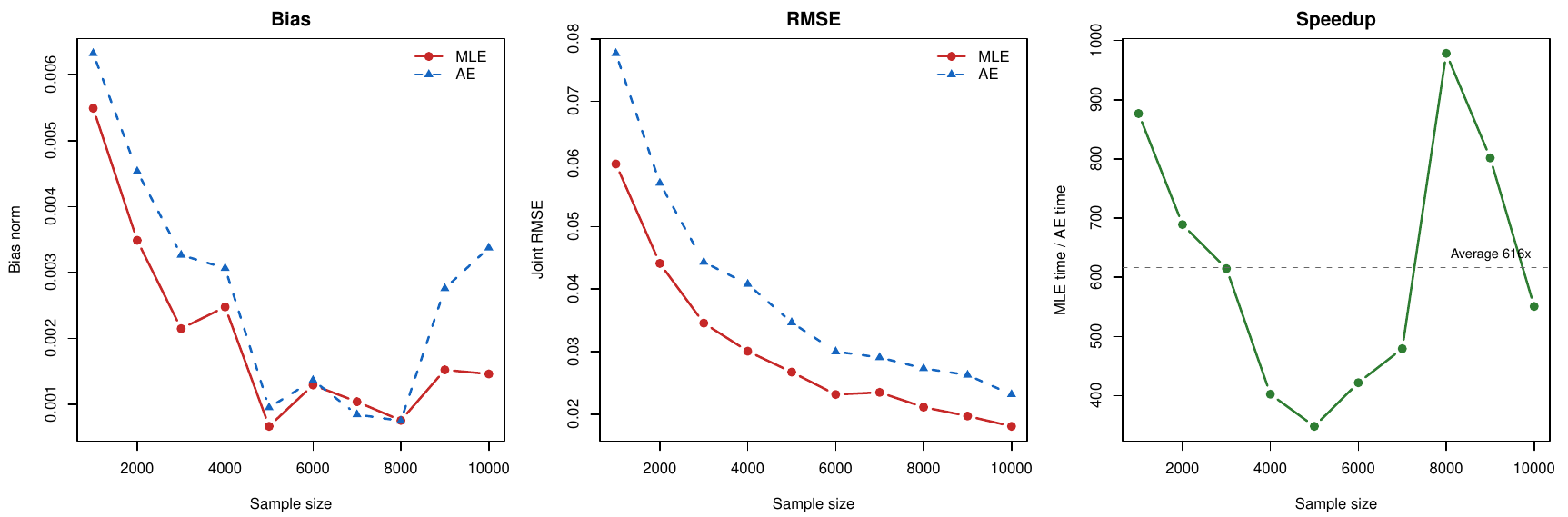}
\caption{Monte Carlo comparison for the Weibull AFT parameters
\((k,\bs\beta^\top)^\top\). Left: Euclidean norm of the bias vector. Center:
joint RMSE. Right: computational speedup, defined as MLE time divided by
AE estimator time.}
\label{fig:weibull-aft-bias-rmse-speedup}
\end{figure}

Figure~\ref{fig:weibull-aft-bias-rmse-speedup} shows that the closed-form
estimator follows the same decreasing bias and RMSE pattern as the MLE, with a
moderate efficiency loss. The average joint RMSE ratio was approximately
\(1.30\). The computational gain is substantial: the AE estimator was on
average about \(616\) times faster than the numerical MLE. This example
illustrates the usefulness of the proposed approach in regression settings,
where repeated likelihood maximization can be costly and closed-form
estimating equations provide a fast noniterative alternative.

\section{Parametric survival trees with closed-form Weibull AFT fits}

We now illustrate how the proposed noniterative estimators can affect the
computational cost of a larger statistical algorithm. The goal is not only to
compare two estimators in isolation, but to show how replacing numerical
likelihood maximization by closed-form auxiliary estimating equations can
change the cost of an inner loop that must be executed many times.

We consider split screening in a parametric survival tree. At a given node, a
large number of candidate splits of the form \(Z_j\leq c\) must be evaluated.
For each candidate split, two Weibull accelerated failure time (AFT) models are
fitted, one in each daughter node, and the split is scored by the resulting
Weibull log-likelihood. This is a typical computational bottleneck in
tree-based survival modelling: before a single split is chosen, the fitting
routine may need to estimate many parametric survival models.

We generate data from a heterogeneous Weibull AFT model. Let
\(Z_1,\ldots,Z_6\) denote candidate splitting variables and let
\(\bs W=(W_1,\ldots,W_4)^\top\) denote covariates entering the AFT model fitted
inside each node. The event time is generated from
\[
\log T_i
=
\mu_{\ell(i)}
+
\bs W_i^\top\bs\beta
+
\frac{1}{k_{\ell(i)}}\varepsilon_i,
\qquad
\varepsilon_i=\log E_i,\quad E_i\sim \operatorname{Exp}(1),
\]
where the latent leaf index \(\ell(i)\) is determined by a two-level tree in
the splitting variables. Thus, the data-generating mechanism contains a true
tree structure, while each terminal region follows a Weibull AFT model with
region-specific intercept and shape parameter.

For each candidate split, we compare two scoring procedures. The first fits
the Weibull AFT model in each daughter node by numerical maximum likelihood.
The second replaces each numerical fit by the closed-form
AE estimator from Section~\ref{sec:aft}, obtained by ordinary least squares on
\(\log T_i\) followed by the residual-variance correction for the Weibull shape
parameter. In both cases, the split is scored by the Weibull log-likelihood
evaluated at the fitted parameters. Therefore, the comparison isolates the
computational effect of replacing repeated likelihood maximization by
noniterative likelihood-derived estimation.

In the experiment, we use \(n=2500\), six candidate splitting variables, four
AFT covariates in each node model, minimum leaf size \(120\), and maximum tree
depth \(2\). Candidate cut points are taken at empirical quantiles
\(0.05,0.10,\ldots,0.95\) of each splitting variable within the current node.
Overall, \(318\) candidate splits are evaluated during tree construction.

\begin{table}[!ht]
\centering
\caption{Weibull AFT split screening in a parametric survival tree.}
\label{tab:weibull-tree-screening}
\begin{tabular}{lrrrr}
\toprule
Method & Internal nodes & Terminal nodes & Splits & Time (s)\\
\midrule
MLE & 3 & 4 & 318 & 6.47\\
AE estimator & 3 & 4 & 318 & 0.24\\
\bottomrule
\end{tabular}
\end{table}

Table~\ref{tab:weibull-tree-screening} shows that both methods selected the
same tree. In particular, the root split was \(Z_2\leq 0.002\) for both the
MLE-based and the AE estimator-based screening procedures, and the
subsequent splits were also identical. The computational cost, however, was
substantially different: the MLE-based tree required \(6.47\) seconds, whereas
the AE estimator-based tree required \(0.24\) seconds, corresponding to a
speedup of approximately \(27\).

\begin{figure}[!ht]
\centering
\includegraphics[width=0.7\textwidth]{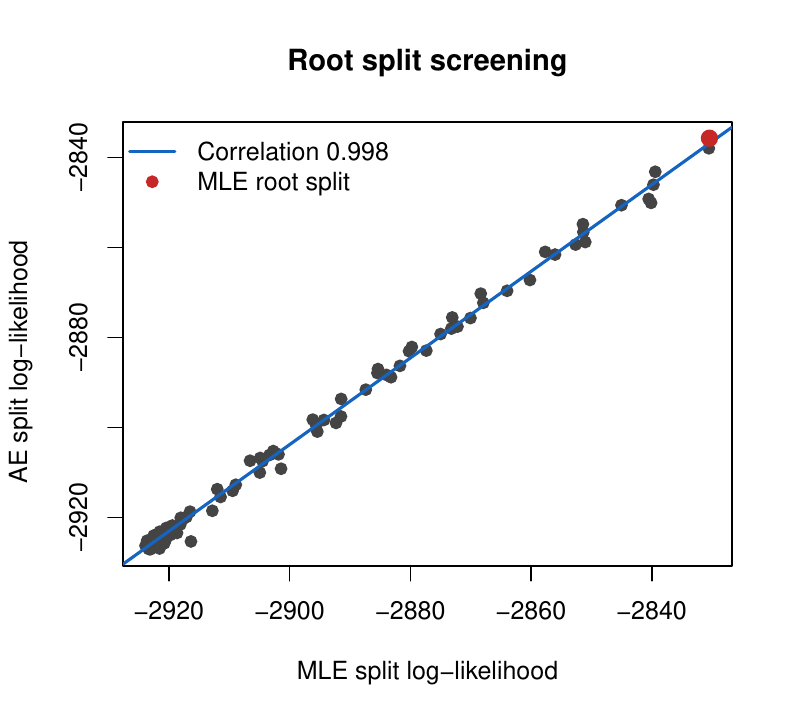}
\caption{Comparison of root-node split scores for the Weibull AFT tree. Each
point corresponds to one candidate split. The horizontal axis reports the
log-likelihood score obtained from numerical MLE fits, and the vertical axis
reports the corresponding score obtained from AE estimator fits.}
\label{fig:weibull-tree-root-scores}
\end{figure}

Figure~\ref{fig:weibull-tree-root-scores} shows that the MLE and
AE estimator split scores are nearly perfectly aligned at the root node;
the empirical correlation between the two sets of scores was \(0.9985\). Thus,
the closed-form estimator preserves the ranking of candidate splits almost
exactly in this experiment, while avoiding hundreds of numerical
optimizations.

The computational implication becomes more pronounced in ensemble settings. A
single tree in this experiment required \(318\) split evaluations. A random
forest with \(500\) trees of comparable size would require approximately
\(159{,}000\) split evaluations. Extrapolating from the observed timings, the
MLE-based procedure would require about \(54\) minutes, whereas the
AE estimator-based procedure would require about \(2\) minutes. This
scaling argument illustrates why noniterative likelihood-derived estimators
are especially attractive inside repeated model-fitting algorithms: they can
preserve likelihood-based split scoring while making parametric survival-tree
and forest procedures substantially more computationally feasible.

\section{Final remarks}

We introduced a likelihood-derived construction of auxiliary estimating
equations that can yield explicit estimators. Each resulting estimator is a
particular just-identified \(Z\)- or GMM estimator; the contribution is the
construction of its estimating function from an auxiliary \(g\), rather than a
new general asymptotic theory. The local result makes the necessary
identification and root-selection conditions explicit and then invokes the
standard estimating-equation limit theory.

The examples show that the auxiliary function need not be the data-generating
density. It may be a normalized extension of the baseline model, as in the
transformed exponential and Beta constructions, or it may be built directly
from unbiased estimating functions, as in the Weibull AFT, Wishart,
zero-inflated count, tail-index and copula examples. This flexibility is both
the usefulness and the limitation of the method: many equations can be encoded,
but explicit solvability, admissibility and efficiency must be checked for the
selected equation. The Godambe result provides an efficiency benchmark, while
the projection result distinguishes harmless interior stabilization from
nonregular boundary behaviour.

The Weibull AFT simulation and survival-tree experiment illustrate the
computational motivation. In the regression simulation, the closed-form
estimator shows the expected decreasing bias and RMSE pattern while avoiding
numerical likelihood maximization. In the survival-tree illustration, replacing
repeated Weibull AFT likelihood maximization by closed-form fits preserved the
selected split structure and substantially reduced computing time. These
results suggest that the proposed estimators are especially useful as
likelihood-derived building blocks inside repeated model-fitting algorithms.

Future work includes systematic selection of auxiliary equations, higher-order
bias correction, robustness analysis, and extensions to censored,
semiparametric, penalized, and high-dimensional models.

\appendix

\section{Algebra for the transformed exponential family}
\label{app:transformed-algebra}

This appendix records the algebra leading to
\eqref{sigmanonweighted}--\eqref{muweighted}. At \(p=1\), differentiation of
\eqref{gweightedexp} with respect to \(\sigma\) gives
\[
\left.\frac{\partial}{\partial\sigma}
\log g(y;\mu,\sigma,p)\right|_{p=1}
=\frac{\mu+1}{\sigma}-\frac1{\sigma+\delta}-\mu T(y).
\]
The derivative with respect to the auxiliary power is
\[
\begin{aligned}
\left.\frac{\partial}{\partial p}
\log g(y;\mu,\sigma,p)\right|_{p=1}
={}&1+\log y
+\frac{T''(y)}{T'(y)}y\log y
+\delta\frac{T'(y)}{1+\delta T(y)}y\log y\\
&-\frac{T'(y)}{T(y)}y\log y
-\mu\sigma T'(y)y\log y
+\mu\frac{T'(y)}{T(y)}y\log y.
\end{aligned}
\]
Both expectations vanish under the normalized embedding, subject to the
common-support regularity stated in Section~\ref{sec:transformed}.

For \(\delta=0\), the empirical \(\sigma\)-equation is
\[
\mu(\sigma^{-1}-A_6)=0,
\]
and hence \(\hat\sigma=A_6^{-1}\). The empirical \(p\)-equation becomes
\[
1+A_1+A_2-A_4+\mu(A_4-\sigma A_5)=0.
\]
Substituting \(\sigma=A_6^{-1}\) and solving for \(\mu\) yields the second
formula in \eqref{sigmanonweighted}.

For \(\delta=1\), the \(\sigma\)-equation is equivalent to
\[
\mu=\frac1{(\sigma+1)(\sigma A_6-1)},
\]
while the \(p\)-equation gives
\[
\mu=\frac{Z}{\sigma A_5-A_4},
\qquad Z=1+A_1+A_2+A_3-A_4.
\]
Equating these expressions produces the quadratic
\[
ZA_6\sigma^2+\{Z(A_6-1)-A_5\}\sigma+(A_4-Z)=0.
\]
The branch in \eqref{sigmaweighted} is the positive root of this equation;
substitution into the second display gives \eqref{muweighted}. The explicit
branch is therefore meaningful only on the sample event where the
discriminant is nonnegative and the resulting \((\hat\mu,\hat\sigma)\) is in
the parameter space, as stated in the main text.

\section{Wishart covariance--log-determinant identity}
\label{app:wishart-identity}

For \(X\sim W_d(\nu,\Sigma)\), temporarily hold \(\Sigma\) fixed. The score
for \(\nu\) can be written as
\[
S_\nu(X)
=\frac12\log|X|-\frac d2\log2-\frac12\log|\Sigma|
-\frac12\psi_d(\nu/2),
\]
where \(\psi_d\) is the multivariate digamma function. The score identity and
\(E_{\nu,\Sigma}(X)=\nu\Sigma\) imply
\[
\Sigma
=\frac{\partial}{\partial\nu}E_{\nu,\Sigma}(X)
=E_{\nu,\Sigma}[X S_\nu(X)]
=\frac12\operatorname{Cov}_{\nu,\Sigma}(X,\log|X|).
\]
Consequently,
\[
E(X\log|X|)-E(X)E(\log|X|)=2\Sigma=\frac{2}{\nu}M,
\qquad M=\nu\Sigma.
\]
Premultiplying by \(M^{-1}\), taking the trace, and centering \(X\) gives
\[
E_{\nu,M}\left[
\operatorname{tr}\{M^{-1}(X-M)\}\log|X|
\right]=\frac{2d}{\nu},
\]
which is the identity used to construct \(\hat\nu_{\mathrm{AE}}\).

\section*{CRediT authorship contribution statement}

Pedro L. Ramos: Conceptualization, Methodology, Formal analysis, Software,
Investigation, Visualization, Writing--original draft, Writing--review and
editing. Eduardo Ramos: Formal analysis, Validation, Writing--review and
editing. Francisco A. Rodrigues: Validation, Writing--review and editing.
Francisco Louzada: Conceptualization, Supervision, Validation,
Writing--review and editing.

\section*{Declaration of competing interest}

The authors declare that they have no known competing financial interests or
personal relationships that could have appeared to influence the work
reported in this paper.

\section*{Data availability}

No external data were used in this study. The numerical results are based on
Monte Carlo experiments generated from the model specifications described in
the paper.

\section*{Acknowledgements}

Eduardo Ramos acknowledges financial support from the S\~ao Paulo State
Research Foundation (FAPESP, grant 2019/27636-9). Francisco A. Rodrigues
acknowledges financial support from CNPq (grant 309266/2019-0). Francisco
Louzada acknowledges financial support from CNPq (grant 301976/2017-1) and
FAPESP (grant 2013/07375-0).

\end{document}